\documentclass[aps,prd,preprint,nofootinbib]{revtex4}
\usepackage{amsfonts}
\usepackage{mathrsfs}
\usepackage{graphicx}
\usepackage{amsmath}
\usepackage{amssymb}
\usepackage{subfigure}
\usepackage{epsfig}
\usepackage{graphicx}
\usepackage{slashed}
\usepackage{color}
\usepackage{float}
\newcommand{\bqa}{\begin{eqnarray}}
\newcommand{\eqa}{\end{eqnarray}}
\newcommand{\beq}{\begin{equation}}
\newcommand{\eeq}{\end{equation}}
\allowdisplaybreaks[1]
\newcommand{\wang}[1]{{\color{blue} #1}}
\usepackage{soul}

\begin{document}

\title{The mass spectrum and wave functions of the $B_c$ system}

\author{\vspace{1cm}Guo-Li Wang$^{1,2}$\footnote[1]{wgl@hbu.edu.cn, corresponding author}, Tianghong Wang$^3$, Qiang Li$^4$,  Chao-Hsi Chang$^{5,6}$\footnote[1]{zhangzx@itp.ac.cn, corresponding author}}

\affiliation{$^1$ Department of Physics, Hebei University, Baoding 071002, China\\
$^2$ Key Laboratory of High-precision Computation and Application of Quantum Field Theory of Hebei Province, Baoding, China\\
$^3$ School of Physics, Harbin Institute of Technology, Harbin 150001, China\\
$^4$ School of Physical Science and Technology, Northwestern Polytechnical University, Xi'an 710072, China\\
$^5$ Institute of Theoretical Physics, Chinese Academy of Science, Beijing 100190, China\\
$^6$ CCAST(World Laboratory), P.O. Box 8730, Beijing 100190, China\vspace{0.6cm}}

\begin{abstract}
\vspace{0.5cm}
The spectrum and relativistic wave functions of $B_c$ system are investigated via solving the complete Salpeter equation. Emphases are put on the study of the partial waves of each $J^P$ state. Our study shows that there are three categories of $J^P$ states. The first category contains $0^-$ and $0^+$ states, which are ${}^1S_0$ dominant state with a small amount of $P$ wave and ${}^3P_0$ dominant state with a small amount of $S$ wave, respectively. The second category includes the natural parity states, such as $1^-$, $2^+$, $3^-$, etc. Taking the $1^-$ state as an example, we study it in two cases. One is the ${}^3S_1$ dominant state with a small amount of $P$ and $D$ waves, and the other is the ${}^3D_1$ dominant state but contains a large amount of $S$ and $P$ wave components. The third category includes the unnatural parity states, such as $1^+$, $2^-$, $3^+$, etc. For the $1^+$ spectrum, the states are grouped into pairs with different radial quantum numbers. Each pair contains two ${}^1P_1-{}^3P_1$ mixing states, and the corresponding mixing angles are calculated by using the relativistic wave functions.

\end{abstract}
 \maketitle
\newpage

\section{Introduction}\label{Sec-1}

In recent years, great progress has been made in the study of hadronic spectra, many new resonances have been found, including the excited double heavy $B_c$ mesons.
The $B_c$ system is unique since it is the only one which carries two different heavy flavors in the Standard Model.
With two different heavy components, its physics is abundant, and draw a lot of attention theoretically in its productions \cite{chang1992,chang127,braaten,cheung,chang4086}, decays \cite{chang3399,kiselev1,chao1997,Colangelo,kiselev2,chang2002,Ivanov,qiao} as well as the mass spectrum \cite{GI,chen,kiselev,fulcher,lattice1,lattice2}. In experiments, people are also very interested in the study of $B_c$ mesons \cite{1998cdf,cdf2006,lhcb2012,lhcb2012-2,lhcb2013}. The latest development is the discovery of the first radial excited pseudoscalar $B_c(2S)$ and vector $B^*_c(2S)$ \cite{bc2s1,bc2s2,bc2s3,bc2sdecay}. An interesting phenomenon, which is predicted in theory and confirmed by experiments, is the mass splitting $M(B^{*}_{c})-M(B_{c})$ being larger than $M(B^{*}_{c}(2S))-M(B_{c}(2S))$ \cite{bc2s2,bc2s3,bc2sdecay}. At the same time, more excited $B_c$ states are expected to be found \cite{qiao2}. All of these motivate us to study the mass spectra of the $B_c$ system.

In this article, we will also focus on the wave functions of the $B_c$ system. We know that some physical states cannot be represented by pure ${}^{2S+1}L_J$ waves, such as the $S-D$ (or $P-F$) mixing states and the ${}^1P_1- {}^3P_1$ (or ${}^1D_2- {}^3D_2$) mixing states.  $\psi(3770)$ is the most typical $S-D$ mixing state, the sizable di-lepton decay width of which indicates that it is not a pure $1{^3D_1}$ wave but contains the $S$ wave component. Because $\psi(3686)$ has a mass close to that of $\psi(3770)$, it is generally believed that these two particles are $2S-1D$ mixing partners,
\begin{equation}\begin{aligned}|\psi(3770)\rangle=|1^3D_1\rangle \cos\theta+|2^3S_1\rangle \sin\theta,\\
|\psi(3686)\rangle=-|1^3D_1\rangle \sin\theta+|2^3S_1\rangle \cos\theta.
\end{aligned}\end{equation}
The $S-D$ mixing is caused by the tensor force \cite{eichten1} and the coupled-channel effects \cite{eichten2,eichten1,heikkila}.
Based on the di-lepton decay widths of $\psi(3770)$, two mixing angles are obtained, $\theta=(12\pm2)^\circ$ and $\theta=-(27\pm2)^\circ$ \cite{rosner} or $\theta=-13^\circ$ and $\theta=26^\circ$ \cite{ding}. Considering some radiative decays and the results of coupled-channel effects, it seems the small angles are favored \cite{rosner,ding}. But the small angles can not explain all the radiative decays \cite{rosner,ding}, and the results of $B^+\to \psi(3770)K^+$ indicate an unexpectedly large mixing angle $|\theta|\approx40^\circ$ \cite{kuiyong},
so it is still an open question about the $S-D$ mixing in the state of $\psi(3770)$.

Similar to the $\psi(3770)$, in the $B^*_c$ system, there may also be the $S-D$ mixing, that is, the wave function of a $B^*_c$ state has both $S$ and $D$ wave components.
So far, there are few studies on the possible $S-D$ mixing of the $B_c$ system, while for the ${}^1P_1- {}^3P_1$ mixing,  there have been many calculations about the mixing angle \cite{davies,godfrey,Ebert,eichten,zhong}. These calculations are done by using different interaction potentials, which could lead to very different mixing angles (see the Table \ref{mass}). Therefore, the two kinds  of mixings of the $B_c$ system need to be carefully studied.

In this article, unlike previous calculations, we use a dynamic method to calculate the mixing angles in the $B_c$ system. In our previous works \cite{Kim:2003ny,Wang:2005qx,0+,chang2005}, based on the $J^P$ quantum number of a meson,  we gave the formula of its wave function, and then solved the  instantaneous Bethe-Salpeter (BS) equation \cite{BS} to obtain the numerical results of the general relativistic wave function. This wave function has a certain $J^P$ quantum number, but contains different partial waves. In this article, we study the different partial waves in the wave function, and calculate their proportions numerically. Unlike $S-D$ mixing and $P-F$ mixing, what we get are $S-P-D$ partial waves in the $1^-$ state and $P-D-F$ partial waves in the $2^+$ state. By solving the complete Salpeter equation \cite{Sal}, which is the instantaneous version of the BS equation, we obtain the mass spectrum of the $B_c$ meson.

This paper is organized as follows. In Sec. II, we give the general form of the relativistic wave function of each $J^P$ state. The partial waves, as well as the corresponding normalized formula, are also presented. In Sec. III, we show our numerical results and draw the conclusions. In Appendix, we present the Bethe-Salpeter equation and show how to get the Salpeter equation.

\section{The relativistic wave functions and the partial waves}\label{Sec-3}
{A meson state is generally represented by ${}^{2S+1}L_J$ or $J^{P(C)}$. The former provides us with a wealth of particle information, such as the spin $S$, the orbital angular momentum $L$, as well as the total angular momentum $J$. The latter can be obtained by using the relations $P=(-1)^{L+1}$ and $C=(-1)^{L+S}$. The ${}^{2S+1}L_J$ form is common used in the literature. However, we must point out that this representation is strictly applicable only for non-relativistic cases, since in a relativistic condition, $S$ and $L$ are no longer good quantum numbers. For example, an orbitally excited $q_1\bar q_2$ state ($q_1$ and $q_2$ are two different quarks) can not be described by $^3P_1$ and $^1P_1$ separately, and the physical state is a mixture of them. The $S-D$ mixing state $\Psi(3770)$ shows us that two different orbital angular momenta $L=0$ and $L=2$ appear in one state.}

We know that, in any case, $J^P$ is a good quantum number which can be used to describe a physical bound state,
and the wave function for a meson should be consistent with it.
We notice that, although BS or Salpeter equation is a relativistic dynamic equation which describes a bound state, the equation itself does not provide the exact kinematic form of the corresponding wave function of a meson. So we construct the relativistic wave function by using the following components: the meson momentum $P$,  the meson mass $M$, the internal relative momentum $q$, the polarization vector $\epsilon_{\mu}$ or tensor $\epsilon_{\mu\nu}$, and the $\gamma$ matrices. Each term in the wave function should have the same quantum number $J^{P}$ as that of the considered meson.

After we get the numerical results of the wave functions, we find the physical states can divide into three categories according to their wave function partial waves:
(1) $0^{-}$ and $0^{+}$ states. They are $^1S_0$ and $^3P_0$ dominated states with a small amount of $P$ and $D$ wave components, respectively. (2) natural $1^{-}$, $2^{+}$, and $3^{-}$ states. Taking $1^{-}$ state as an example, it can be a $^3S_1$ dominated state with a small amount of $P$ and $D$ wave components, or it can be a $^3D_1$ dominated state, but contains a large amount of $S$ and $P$ wave components. (3) unnatural $1^{+}$ and $2^{-}$ states. They are always mixing states. $1^{+}$ is a $^3P_1-{}^1P_1$ mixture, and $2^{-}$ is the mixing of $^3D_2$ and ${}^1D_2$. We also point out that, $1^{+}$ is not a pure $P$ wave, but contains a small amount of $D$ wave component, and $2^{-}$ is a $D$ wave dominant state with a small mount of $F$ wave.

\subsection{$0^{-}$ and $0^{+}$ states}

\subsubsection{The wave function of a $0^{-}$ state and its partial waves}
With the instantaneous approximation ($P\cdot q=0$), the general relativistic wave function for a pseudoscalar can be written as \cite{Kim:2003ny},
\begin{equation}
\begin{aligned}\label{pseudowave}
\varphi_{_P}^{0^-}({q}_{_\bot})=\displaystyle
\left(a^{}_1 M+a^{}_2\not\!P+a_3\not\!{q}_{_\bot}
+a_4\frac{\not\!{q}_{_{\bot}}\not\!P}{M}\right)\gamma^{5},
\end{aligned}
\end{equation}
where $M$ and $P$ are the mass and momentum of the meson, respectively; $q$ is the relative momentum between quark and antiquark inside the meson;  ${q}_{_{\bot}}\equiv{q}_{_{P\bot}}=q-\frac{P\cdot q}{M^2}P$; $a_i~(i=1,2,3,4)$ is the radial part of the wave function, and it is a function of $-{q}_{_{\bot}}^2=\vec{q}^2$. The four $a_i s$ are not all independent. According to the constrain conditions in Eq. (\ref{const}), two of them are independent, then the wave function for the $0^{-}$ state is written as
\begin{equation}\label{pseudowave1}
\begin{aligned}
\varphi_{_P}^{0^{-}}({q}_{_{\bot}})=\displaystyle
M\left(a^{}_1+a^{}_2 \frac{\not\!P}{M}-
{a_1} x_{-} \not\!{q}_{_{\bot}}
+{a_2} x_{+} \frac{\not\!{q}_{_{\bot}}\not\!P}{M}\right)\gamma^{5},
\end{aligned}
\end{equation}
where
$$x_{+}=\frac{\omega_{1}+\omega_{2}}{m_1\omega_{2}+m_2\omega_{1}},~~ x_{-}=\frac{\omega_{1}-\omega_{2}}{m_1\omega_{2}+m_2\omega_{1}},$$
with $\omega_1=\sqrt{m^2_1-{q}^2_{_{\bot}}}$, $\omega_2=\sqrt{m^2_2-{q}^2_{_{\bot}}}$. $m_1$ and $m_2$ are the constituent quark masses of quark $1$ and antiquark $2$, respectively.

{To see the parity of the wave function, we make the following transformation,
$$\varphi_{_P}({q})=\eta_{_P}\gamma_0\varphi_{_{P'}}({q'})\gamma_0,$$
where $P'=(P_0,-\vec P)$ and $q'=(q_0,-\vec q)$. Then we can check that every term in Eq.(\ref{pseudowave}) has the negative parity ($\eta_{_P}=-$). When $m_1=m_2$, we can further consider the charge conjugation transformation,
$$\varphi_{_P}({q})=\eta_{_C}C\varphi^{T}_{_P}({-q})C^{-1},$$
where $T$ is the rotation transform, $C=\gamma_2\gamma_0$ with $C\gamma^T_5C^{-1}=\gamma_5$ and $C\gamma^T_{\mu}C^{-1}=-\gamma_{\mu}$.
We can see that the terms including $a_1$, $a_2$ and $a_4$ have positive $C$ parity ($\eta_{_C}=+$), while the $a_3$ term has  negative $C$ parity. } This may seem wrong at first glance. But after applying the constrain condition in Eq.(\ref{const}), we can see that the $a_3$ term in Eq.(\ref{pseudowave1}) will disappear because of $\omega^{}_1-\omega^{}_2=0$. So the wave function of $0^-$ state automatically becomes to the one of the $0^{-+}$ state if $m_1=m_2$. The same thing happens to all the states with the natural parity $(-1)^J=$ $0^+$, $1^-$, $2^+$, $3^-$, etc., which shows that our method is correct and self-consistent.

Now there are two unknown independent radial wave functions, $a_1$ and $a_2$. The normalization formula is
\begin{equation}\label{0-nor}
\int\frac{d{\vec q}}{(2\pi)^3}\frac{8M\omega_{1}\omega_{2}a_1
a_2 }{(\omega_{1}m_{2}+\omega_{2}m_{1})}
=1.
\end{equation}
With the wave function Eq.(\ref{pseudowave1}) of the $0^{-}$ state as an input, we can solve the coupled equations, i.e. Eq.(\ref{posi}) and Eq.(\ref{nega}) and obtain the numerical results of $a_1$ and $a_2$ \cite{Kim:2003ny}. As an eigenvalue problem, the mass spectrum of $0^{-}$ state is also obtained at the same time.

{
Although the wave function in Eq.(\ref{pseudowave}) is a general expression for a $0^-$ state, it is not a pure ${}^1S_0$ wave, but contains a $P$ wave component. Other higher partial waves such as the $D$ wave are suppressed by the instantaneous approximation and ignored here. Similar conclusions happen to other $J^P$ states, and will not be mentioned. To see the details of partial waves, we rewrite the wave function Eq.(\ref{pseudowave}) in terms of the spherical harmonics,
\begin{equation}
\begin{aligned}
\varphi_{_P}^{0^-}({q}_{_\bot})=\displaystyle
 \sqrt{4\pi}\left[ MY_{00}\left(a^{}_1 +a^{}_2\gamma^0 \right)-\frac{|\vec q|}{\sqrt{3}}( Y_{1-1}\gamma^+ +Y_{11}\gamma^- -Y_{10}\gamma^3)(a_3
+a_4\gamma^0)\right]\gamma^{5},
\end{aligned}
\end{equation}
where $\gamma^{\pm}=\mp\frac{1}{\sqrt{2}}(\gamma^1\pm \gamma^2)$, and $Y_{Lm}$ is the spherical harmonic function. We can see that, the $a_1$ and $a_2$ terms are $S$ waves, while $a_3$ and $a_4$ terms are $P$ waves. Since we have shown that all terms have negative parity, the relation $P=(-1)^{L+1}$ working for the dominant $a_1$ and $a_2$ terms which survive in the non-relativistic limit, is no longer applicable to the small relativistic correction $a_3$ and $a_4$ terms which will disappear in the non-relativistic condition. Same conclusion is suitable for all other states, and will not be mentioned again. In the following part, instead of showing the details of partial waves in terms of the spherical harmonics, we give a simple way to roughly distinguish the partial waves, that is to count the number of $q_{_\perp}$: zero-$q_{_\perp}$ term is $S$ wave, one-$q_{_\perp}$ term is $P$ wave, two-$q_{_\perp}$ term is $D$ wave, three-$q_{_\perp}$ term is $F$ wave, and so on.  }

In Eq.(\ref{pseudowave}), if we delete the $P$ wave components, then the wave function becomes to the familiar one
\begin{equation}
\varphi_{_P}^{{}^1S_0}({q}_{_\bot})=\displaystyle
\left(a^{}_1 M+a^{}_2\not\!P\right)\gamma^{5},
\end{equation}
which only contains the pure ${}^1S_0$ wave component. In this case,
the normalization condition is
\begin{equation}\label{0-pure}
\int\frac{d{\vec q}}{(2\pi)^3}\frac{2Ma_1
a_2 (\omega_{1}m_{2}+\omega_{2}m_{1}) }{\omega_{1}\omega_{2}}
=1.
\end{equation}
Combining Eq.(\ref{0-nor}) and Eq.(\ref{0-pure}), the contributions of $S$ and $P$ partial waves can be separated.

\subsubsection{The wave function of a $0^{+}$ state and its partial waves}
The general relativistic wave function of a
 $0^{+}$ state is written as \cite{0+},
\begin{equation}\label{0+}
\varphi^{0^{+}}_{_P}({q}_{_{\bot}})=
b_1~{\not\!q_{_\perp}}+b_2~\frac{{\not\!P}
{\not\!q}_{_\perp}}{M} +b_3~M+b_4~{\not\!P}.
\end{equation}
The constraint conditions indicate
$$b_3=\frac{b_1~q_{_\perp}^2 x_+}
{M},~~~
b_4=\frac{b_2~q_{_\perp}^2 x_{-}}
{M}.$$
And the normalization
condition is
 \begin{equation}\label{0+nor}
\int \frac{d{\vec q}}{(2\pi)^3}\frac{8
\omega_1\omega_2{\vec q}^2 b_1 b_2}{M(m_1\omega_2+m_2\omega_1)}=1.
 \end{equation}

If we only consider the  pure ${}^3P_0$ wave component, the $b_3$ and $b_4$ terms, which represent the $S$ wave, will disappear, then Eq.(\ref{0+}) becomes
\begin{equation}
\varphi^{{}^3P_0}_{_P}({q}_{_{\bot}})=
b_1~{\not\!q_{_\perp}}+b_2~\frac{{\not\!P}
{\not\!q}_{_\perp}}{M}.
\end{equation}
And the corresponding normalization condition is
 \begin{equation}\label{0+pure}
\int \frac{d{\vec q}}{(2\pi)^3}\frac{2
{\vec q}^2 b_1 b_2 (m_1\omega_2+m_2\omega_1)}{M\omega_1\omega_2}=1.
 \end{equation}

\subsection{$1^{-}$, $2^{+}$ and $3^{-}$ states}

\subsubsection{The wave function of a $1^{-}$ state and its partial waves}
The relativistic wave function for a
vector state is written as \cite{Wang:2005qx,1-solve},
\begin{equation}\begin{aligned}\label{vecwave}\varphi^{1^{-}}_{_P}({q}_{_{\bot}})&=
{\epsilon}\cdot{q}_{_{\bot}}
\left[c_1+\frac{\not\!P}{M}~c_2+
\frac{{\not\!{q}_{_{\bot}}}}{M}~c_3+\frac{{\not\!P}
{\not\!{q}_{_{\bot}}}}{M^2}~c_4\right]+
M{\not\!\epsilon}~c_5
\\&+
{\not\!\epsilon}{\not\!P}~c_6+
({\not\!{q}_{_{\bot}}}{\not\!\epsilon}-
{\epsilon}\cdot{q}_{_{\bot}})
~c_7+\frac{1}{M}({\not\!P}{\not\!\epsilon}
{\not\!{q}_{_{\bot}}}-{\not\!P}{\epsilon}\cdot{q}_{_{\bot}})
~c_8,
\end{aligned}
\end{equation}
where ${\epsilon}$ is the polarization
vector of the meson. According to Eq.(\ref{const}), only  four radial wave functions are independent. Here we choose $c_3$, $c_4$, $c_5$ and $c_6$, and then get
$$c_1=\frac{M^2 c_5+{q}_{_{\bot}}^2 c_3}{M} x_+,~~c_2=\frac{{q}_{_{\bot}}^2 c_4-M^2 c_6}{M} x_-,~~c_7=-M c_5 x_{-},~~c_8=-M c_6 x_{+}.$$

If only the pure ${}^3S_1$ wave is considered, Eq.(\ref{vecwave})  becomes
\begin{equation}
\varphi^{^3S_1}_{_P}({q}_{_{\bot}})=M{\not\!\epsilon}~c_5+{\not\!P}{\not\!\epsilon}~c_6,
\end{equation}
and the normalization condition is
\begin{equation}\label{1-Swave}
-\int\frac{d{\vec q}}{(2\pi)^3}\frac{2Mc_5c_6(\omega_{1}m_{2}+\omega_{2}m_{1})}
{\omega_{1}\omega_{2}} =1.
\end{equation}
While for a pure ${}^3D_1$ state, the wave function is
\begin{equation}\varphi^{^3D_1}_{_P}({q}_{_{\bot}})={\epsilon}\cdot{q}_{_{\bot}}
\left(
\frac{{\not\!{q}_{_{\bot}}}}{M}~c_3+\frac{{\not\!P}
{\not\!{q}_{_{\bot}}}}{M^2}~c_4\right),\end{equation}
with the normalization condition
\begin{equation}\label{1-Dwave}
\int\frac{d{\vec q}}{(2\pi)^3}\frac{2c_3c_4{\vec q}^4 (\omega_{1}m_{2}+\omega_{2}m_{1})}
{3M^3\omega_{1}\omega_{2}} =1.
\end{equation}
Other terms including $c_1$, $c_2$, $c_7$ and $c_8$ are $P$ waves, so the complete $1^{-}$ wave function includes $S$, $P$ and $D$ waves. The relativistic dynamic BS or Salpeter equation will determine the numerical values of the $S$, $P$ and $D$ partial waves.

The normalization formula for the complete wave function of the $1^{-}$ state is
\begin{equation}\label{1-nor}
\int\frac{d{\vec q}}{(2\pi)^3}\frac{8 M\omega_{1}\omega_{2}}{3(\omega_{1}m_{2}+\omega_{2}m_{1})}
\left[ -3c_5c_6+\frac{{\vec q}^2}{M^2}(-c_4c_5+c_3 c_6+c_3c_4\frac{{\vec q}^2}{M^2}) \right]=1.
\end{equation}
Because the whole wave function is $S+P+D$, this normalization formula represents $S^2+P^2+D^2+2(S\cdot P+S\cdot D+P\cdot D)$. The expressions of pure $S^2$ and $D^2$ are already shown in Eq.(\ref{1-Swave}) and Eq.(\ref{1-Dwave}). By subtracting the left part of Eq.(\ref{1-nor}) from the left parts of Eq.(\ref{1-Swave}) and Eq.(\ref{1-Dwave}), we get $P^2+2(S\cdot P+S\cdot D+P\cdot D)$. Using these relations and the numerical solution of the Salpeter equation, we can obtain the contributions of $S$, $P$ and $D$ partial waves.

\subsubsection{The wave function of a $2^{+}$ state and its partial waves}
The relativistic wave function of a
$2^{+}$ state can be written as \cite{2+},
\begin{equation}
\begin{aligned}
\varphi^{2^{+}}_{_P}({q}_{_\bot})&=
{\epsilon}_{\mu\nu}{q_{_\perp}^{\mu}}
\left\{{q_{_\perp}^{\nu}}\left[d_1+\frac{\not\!P}{M}d_2+
\frac{{\not\!q}_{_\perp}}{M}d_3+\frac{{\not\!P}
{\not\!q}_{_\perp}}{M^2} d_4\right]\right.\\
&\left.+{\gamma^{\nu}}\left(Md_5+ {\not\!P}d_6\right)+
({\not\!q}_{_\perp}\gamma^{\nu}-{q}_{_\perp}^{\nu}) d_7+\frac{(\gamma^{\nu}{\not\!q}_{_\perp}-{q}_{_\perp}^{\nu}){\not\!P}}{M}
d_8\right\},\label{2+wave}
\end{aligned}
\end{equation}
where ${\epsilon}_{\mu\nu}$ is the polarization tensor of
the $2^{+}$ state. From the constraint conditions, we get
$$d_1={\frac{\left(q_{\perp}^2 d_3 +M^2 d_5
\right)x_{+}}{M}},~~d_2={\frac{\left(q_{\perp}^2 d_4 -M^2 d_6\right)
x_{-}}{M}},~~d_7=-{d_5Mx_{-}}
,~~d_8=-{d_6Mx_+}.$$

If only pure ${}^3P_2$ wave is considered, Eq.(\ref{2+wave}) becomes
\begin{equation}
\varphi^{^3P_2}_{_P}({q}_{_\bot})={\epsilon}_{\mu\nu}{q_{_\perp}^{\mu}} {\gamma^{\nu}}(Md_5+ {\not\!P}d_6),
\end{equation}
with the normalization condition
\begin{equation}\label{2+P}
-\int\frac{d{\vec q}}{(2\pi)^3}\frac{2d_5d_6{\vec q}^2 M (\omega_{1}m_{2}+\omega_{2}m_{1})}
{3\omega_{1}\omega_{2}} =1.
\end{equation}
While for a pure ${}^3F_2$ state, its wave function is
\begin{equation}
\varphi^{^3F_2}_{_P}({q}_{_\bot})={\epsilon}_{\mu\nu}{q_{_\perp}^{\mu}}
{q_{_\perp}^{\nu}}\left[
\frac{{\not\!q}_{_\perp}}{M}d_3+\frac{{\not\!P}
{\not\!q}_{_\perp}}{M^2} d_4\right],
\end{equation}
and its normalization is
\begin{equation}\label{2+F}
\int\frac{d{\vec q}}{(2\pi)^3}\frac{4d_3d_4{\vec q}^6 (\omega_{1}m_{2}+\omega_{2}m_{1})}
{15M^3\omega_{1}\omega_{2}} =1.
\end{equation}
Other terms with radial wave functions $d_1$, $d_2$, $d_7$ and $d_8$ are $D$ waves, so the complete $2^{+}$ wave function includes $P$, $D$ and $F$ wave components.

The normalization formula for the complete $2^{+}$ state is
\begin{equation}\label{2+nor}
\int\frac{d{\vec q}}{(2\pi)^3}\frac{8 M\omega_{1}\omega_{2}{\vec
q}^2}{15(\omega_{1}m_{2}+\omega_{2}m_{1})}
\left[ -5d_5d_6+\frac{2{\vec q}^2}{M^2}(-d_4d_5+d_3 d_6+d_3d_4\frac{{\vec q}^2}{M^2}) \right]=1.
\end{equation}
Similar to the case of $1^-$ state, the complete wave function of $2^+$ state is $P+D+F$, the normalization formula Eq.(\ref{2+nor}) is $P^2+D^2+F^2+2(P\cdot D+P\cdot F+D\cdot F)$. The expressions of pure $P^2$ and $F^2$ are shown in Eq.(\ref{2+P}) and Eq.(\ref{2+F}), respectively. From Eq.(\ref{2+nor}), we can obtain $D^2+2(P\cdot D+P\cdot F+D\cdot F)$, and finally get the contributions of $P$, $D$ and $F$ waves in a $2^+$ state.

\subsubsection{The wave function of a $3^{-}$ state and its partial waves}
The general wave function for a $3^{-}$ state is given as follows \cite{2--3-},
\begin{equation}
\begin{aligned}\label{3-wave}
\varphi^{3^{-}}_{_P}(q_{_\perp})& = \epsilon_{\mu\nu\alpha}q^\mu_{_\perp} q^\nu_{_\perp}\left[q_{_\perp}^\alpha\left(e_1
 + \frac{\slashed P}{M}e_2 + \frac{{\not\!q}_{_\perp}}{M}e_3 + \frac{\slashed P{\not\!q}_{_\perp}}{M^2}e_4\right)
 +M\gamma^\alpha \left(e_5 + \frac{\slashed P}{M} e_6 \right.\right.\\
&\left.\left.+ \frac{{\not\!q}_{_\perp}}{M}e_7 + \frac{\slashed P{\not\!q}_{_\perp}}{M^2} e_8\right)\right],
\end{aligned}
\end{equation}
where $\epsilon^{\mu\nu\alpha}$ is the third-order polarization tensor of the meson. We choose $e_3\sim e_6$ as the independent radial wave functions, then
$$
e_1=-e_3\frac{\vec q^2 x_{+}}{M}+e_5 M(x_{+}-x_{-}),~~e_2=-e_4\frac{\vec q^2 x_{+}}{M}+e_6 M(x_{+}-x_{-}),~~
e_7=e_5 Mx_{-},~~e_8=e_6 M x_{+}.
$$

We note that there are $D$, $F$ and $G$ partial waves in the wave function Eq.(\ref{3-wave}). If we only consider the pure $^3D_3$ wave, the wave function becomes
\begin{equation}
\begin{aligned}
\varphi^{^3D_3}_{_P}(q_{_\perp})= \epsilon_{\mu\nu\alpha}q^\mu_{_\perp} q^\nu_{_\perp}\gamma^\alpha \left(Me_5 +{\slashed P} e_6 \right),
\end{aligned}
\end{equation}
with the normalization condition
\begin{equation}\label{3-D}
\int\frac{d{\vec q}}{(2\pi)^3}\frac{4e_5e_6{\vec q}^4 M (\omega_{1}m_{2}+\omega_{2}m_{1})}
{15\omega_{1}\omega_{2}} =1.
\end{equation}
While for a pure ${}^3G_3$ state, the wave function is
\begin{equation}
\varphi^{^3G_3}_{_P}(q_{_\perp})=\epsilon_{\mu\nu\alpha}q^\mu_{_\perp} q^\nu_{_\perp}q_{_\perp}^\alpha\left( \frac{{\not\!q}_{_\perp}}{M}e_3 + \frac{\slashed P{\not\!q}_{_\perp}}{M^2} e_4\right),
\end{equation}
and the normalization condition is
\begin{equation}\label{3-G}
-\int\frac{d{\vec q}}{(2\pi)^3}\frac{4e_3e_4{\vec q}^8 (\omega_{1}m_{2}+\omega_{2}m_{1})}
{35M^3\omega_{1}\omega_{2}} =1.
\end{equation}
Other terms with radial wave functions $e_1$, $e_2$, $e_7$ and $e_8$ are $F$ waves.

The normalization condition of the complete wave function is
\begin{equation}\label{3-nor}
\int\frac{d\vec q}{(2\pi)^3}\frac{16\omega_1\omega_2\vec q^4}{105M(m_1\omega_2+m_2\omega_1)}\left(-\frac{3\vec q^4}{M^2}e_3e_4-3\vec q^2e_3e_6 + 3\vec q^2e_4e_5+7M^2e_5e_6\right) = 1.
\end{equation}
Since we know the expressions of pure $D^2$ component in Eq.(\ref{3-D}) and pure $G^2$ in Eq.(\ref{3-G}) and whole expression $(D+F+G)^2$ in Eq.(\ref{3-nor}), we can obtain the contributions of $D$, $F$ and $G$ wave components in a $3^-$ state.

\subsection{$1^{+}$ and $2^{-}$ states}

\subsubsection{The wave function of a $1^{+}$ state and its partial waves}
The general relativistic wave function for a $1^+$ state can be written as \cite{1+},
\begin{equation}
\begin{aligned}\label{1+wave}
\varphi_{_P}^{1^+}({q}_{_\bot})&=\displaystyle \epsilon \cdot q^{}_{_\bot}\left(f^{}_1+f^{}_2\frac{\not\!P}{M}+f_3\frac{\not\!q^{}_{_\bot}}{M}
+f_4\frac{\not\!P\not\!q^{}_{_\bot}}{M^2}\right)\gamma^{5}
\\
&+\displaystyle\frac{i\varepsilon^{}_{\mu\nu\rho\sigma}
\gamma^{\mu}_{}P^{\nu}_{}q^{\rho}_{_\bot}\epsilon^{\sigma}}{M}\left(g^{}_1+g^{}_2\frac{\not\!P}{M}+g_3\frac{\not\!q^{}_{_\bot}}{M}
+g_4\frac{\not\!P\not\!q^{}_{_\bot}}{M^2}\right),
\end{aligned}
\end{equation}
where $\epsilon$ is the polarization vector of the $1^+$ state; $\varepsilon^{}_{\nu\lambda\rho\sigma}$ is the {\it Levi-Civita} symbol. Because of the constraints, only four radial wave functions are independent
\begin{equation}
\begin{aligned}\label{1+wave1}
\varphi^{1^+}_{_P}({q}_{_\bot})&=\displaystyle \epsilon\cdot q^{}_{_\bot}\left(f^{}_1+f^{}_2\frac{\not\!P}{M}-{f_1 x_{-}\not\!q^{}_{_\bot}}
-f^{}_2x_{+}\frac{\not\!P\not\!q^{}_{_\bot}}{{M}}\right)\gamma^{5}
\\
&+\displaystyle\frac{i\varepsilon^{}_{\mu\nu\rho\sigma}
\gamma^{\mu}_{}P^{\nu}_{}q^{\rho}_{_\bot}\epsilon^{\sigma}}{M}\left(g^{}_1+g^{}_2\frac{\not\!P}{M}+{g_1 x_{-}\not\!q^{}_{_\bot}}
+g^{}_2x_{+}\frac{\not\!P\not\!q^{}_{_\bot}}{{M}}\right).
\end{aligned}
\end{equation}

One may note that, we use two kinds of symbols $f_i$ and $g_i$ to represent the radial wave functions. The reason is that for a $m_1=m_2$ system, the $f_i$ terms, which have negative $C$ parity ($f_3$ term disappears because $x_{-}=0$), are $1^{+-}$ (${}^1P_1$) states, while the $g_i$ terms, which have positive $C$ parity ($g_3$ term disappears),  are $1^{++}$ (${}^3P_1$) states. The constrain condition does not mix the $f_i$ and $g_i$ terms because they have different $C$ parities (see Eq.(\ref{1+wave1})). So the general wave function of the $1^+$ state with $m_1\neq m_2$ is a ${}^1P_1-{}^3P_1$ mixture.

Similarly, the numerical values of the radial wave functions are determined by the dynamic BS or Salpeter equation. If we consider a quarkonium, all the solutions of  Eq.(\ref{1+wave1}) will automatically become $1^{+-}$ or $1^{++}$ states, that is the solutions with definite $C$ parities. For example, the first solution is a pure $1^{++}$ state with $f_i=0$, and the second one is a pure $1^{+-}$ state with $g_i=0$. When a non-equal mass system is considered, all the solutions are mixed states of ${}^1P_1$ and ${}^3P_1$, that is they are all ${}^1P_1 -{}^3P_1$ mixtures with $f_i\neq 0$ and $g_i \neq 0$.

The normalization condition for the $1^+$ wave function is
 \begin{equation}\label{1+nor}
\int \frac{d{\vec q}}{(2\pi)^3}\frac{8
\omega_1\omega_2{\vec q}^2}{3M(m_1\omega_2+m_2\omega_1)}(f_1f_2-2g_1g_2)\equiv cos^2\theta+sin^2\theta=1.
 \end{equation}
Since $f_i$ and $g_i$ terms are ${}^1P_1$ and ${}^3P_1$ states, respectively, we can define their mixing angle $\theta$ (see the last formula in Eq.(\ref{1+nor})).

It must be pointed out that we previously called the $f_i$ and $g_i$ terms as ${}^1P_1$ and ${}^3P_1$ states, which is not very strict. They actually contain a small amount of $D$ wave components. The terms $f_1$, $f_2$, $g_1$ and $g_2$ are pure $P$ waves and have dominant contribution, while the terms $f_3$, $f_4$, $g_3$ and $g_4$ are $D$ waves with small contribution. If we ignore the small $D$ wave terms in Eq.(\ref{1+wave}),
the normalization condition of the pure $P$ wave part is
 \begin{equation}\label{1+pure}
\int \frac{d{\vec q}}{(2\pi)^3}\frac{2(m_1\omega_2+m_2\omega_1)
{\vec q}^2}{3M\omega_1\omega_2}(f_1f_2-2g_1g_2)\equiv cos^2\varphi+sin^2\varphi=1.
 \end{equation}
Comparing Eq.(\ref{1+nor}) and Eq.(\ref{1+pure}), we can obtain the contributions of $P$ and $D$ partial waves as well as the mixing angles $\theta$ and $\varphi$.

\subsubsection{The wave function of a $2^{-}$ state and its partial waves}
The general wave function of a $2^{-}$ state with the polarization tensor $\epsilon_{\mu\nu}$ is written as
\begin{equation}
\label{2-wave}
\begin{aligned}
\varphi^{2^{-}}_{_P}(q_{_\perp})&=\epsilon_{\mu\nu}q_{_\perp}^\mu
q_{_\perp}^\nu\left(h_1 + \frac{\slashed P}{M} h_2 + \frac{{\not\!q}_{_\perp}}{M}h_3
 + \frac{\slashed P{\not\!q}_{_\perp}}{M^2} h_4
\right)\gamma^5
\\ &+\frac{ i\varepsilon^{}_{\mu\nu\alpha\beta}\gamma^\mu{P^\nu}q_{_\perp}^\alpha\epsilon^{\beta\delta}q_{{_{\perp \delta}}}}{M} \left(i_1 + \frac{\slashed P}{M} i_2 + \frac{{\not\!q}_{_\perp}}{M}i_3 + \frac{\slashed P{\not\!q}_{_\perp}}{M^2}i_4\right).
\end{aligned}
\end{equation}
The first four terms in Eq.(\ref{2-wave}), whose radial parts are labeled as $h_i$, are $^1D_2$ waves ($J^{PC}=2^{-+}$ in the condition of $m_1=m_2$), and the last  four terms are $^3D_2$ waves ($2^{--}$ when $m_1=m_2$), so the general wave function for a $2^-$ state is a ${}^1D_2 - {}^3D_2$ mixture.

After applying the constrain condition, we obtain
$$h_3=-h_1Mx_-,~h_4=-h_2Mx_+,~i_3=i_1Mx_-,~i_4=i_2Mx_+.$$
It should be noted that the constrain condition does not mix $h_i$s and $i_i$s, because they have different $C$ parities when $m_1=m_2$. The normalization condition for the whole $2^-$ wave function is
 \begin{equation}\label{2-nor}
\int \frac{d{\vec q}}{(2\pi)^3}\frac{8
\omega_1\omega_2{\vec q}^4}{15M(m_1\omega_2+m_2\omega_1)}(2h_1h_2-3i_1i_2)\equiv cos^2\theta+sin^2\theta=1,
 \end{equation}
where in the second equation, we have defined the mixing angle $\theta$ between ${}^1 D_2$ and ${}^3D_2$.

Similar to the case of $1^+$, for the quarkonia, the wave functions of the $2^{-+}$ and $2^{--}$ states automatically decouple from each other. That is, the first solution, third solution, etc, are pure $2^{--}$ states with $h_i=0$, and the second, forth solution, etc, are pure $2^{-+}$ states with $i_i=0$. When a non-equal mass system is considered, all the solutions are mixed states of ${}^1 D_2$ and ${}^3D_2$ with $h_i\neq 0$ and $i_i \neq 0$. Using the numerical result of the normalization condition Eq.(\ref{2-nor}), we can obtain the value of mixing angle $\theta$.

We should also point out that only $h_1$ and $h_2$ terms are pure ${}^1 D_2$ waves and they will provide the dominant contribution, while $h_3$ and $h_4$ terms are $F$ waves and they have small contributions. Similarly, the $i_1$ and $i_2$ terms are pure ${}^3 D_2$ waves with large contributions, while the $i_3$ and $i_4$ terms are $F$ waves with small contributions.
If we ignore the small $F$ waves, then the normalization condition for the pure $D$ wave is
 \begin{equation}\label{2-pure}
\int \frac{d{\vec q}}{(2\pi)^3}\frac{2
{\vec q}^4(m_1\omega_2+m_2\omega_1)}{15M\omega_1\omega_2}(2h_1h_2-3i_1i_2)\equiv cos^2\varphi+sin^2\varphi=1.
 \end{equation}

\section{Numerical results and discussions}

The wave functions we have constructed already have the relativistic forms, so in order to avoid double counting of relativistic corrections, we only need the simple non-relativistic interaction potential when solving the complete Salpeter equation.  The potential we choose has the form of the Coulomb vector part plus a linear confinement part and a free parameter $V_0$.
In our calculation, we use the well-fitted constituent quark masses $m_c=1.62\ {\rm GeV}$, $m_b=4.96\ {\rm GeV}$ and other model parameters which can be found in Refs.\cite{fuhuifeng,Wang:2013lpa}.

\subsection{Mass spectra}

Our result of the mass spectrum is shown in Table \ref{mass}. In order to adjust the ground state eigenvalue of each $J^P$ state, we choose different $V_0$ values, which correspond to different $J^P$ states that we use different wave functions. Therefore, the ground state eigenvalues of each $J^P$ state are input values, and are marked with `input' in Table \ref{mass}, in this case what we predict are actually the mass splittings.

In Table \ref{mass}, as a comparison, we also list the results of some other theoretical models, as well as the experimental data. It can be seen that for the theoretical masses of the low excited states, our results are in good agreement with those of other theoretical models and the experimental values. The mass splittings $M(B^*_c(2S)^+)-M(B^{*+}_c)$ and $M(B_c(2S)^+)-M(B_c^+)$ are particularly noteworthy. One can see the former is slightly smaller than the latter, and all the theoretical predictions including ours are consist with experimental data. For the highly excited states, there is no experimental value at present, and the theoretical calculation results are in good agreement with each other. This is due to the fact that the $B_c$ system composed of double heavy quarks is very heavy, then the relativistic correction is relatively small, so that the theoretical results of different models are not much different.

\begin{table*}[hbt]
\caption{Mass spectra and mixing angles of the $B_c$ system.}\label{mass}
\begin{tabular*}{\textwidth}{@{}c@{\extracolsep{\fill}}ccccccccc}
\hline \hline
$n~^{2S+1}L_J$&$J^P$&\rm{ours}&\cite{godfrey}&\cite{Ebert}&\cite{eichten}&\cite{zhong}&\cite{lattice}&Exp\\ \hline
%{\phantom{\Large{l}}}\raisebox{+.2cm}{\phantom{\Large{j}}}
$1~^1S_0$ &$0^-$&6277~({\scriptsize{input}})&6271&6272&6275&6271~({\scriptsize{input}})&6276&6274.9$\pm$0.8~\cite{pdg}
\\
$1~^3S_1$ &$1^-$&6332~({\scriptsize{input}})&6338&6333&6329&6326~({\scriptsize{input}})&6331&6333~\cite{bc2s2}
\\
$2~^1S_0$ &$0^-$&6867&6855&6842&6867&6871~({\scriptsize{input}})&&6871.6$\pm$1.1~\cite{pdg}
\\
$2~^3S_1$ &$1^-$&6911&6887&6882&6898&6890&&6900.1~\cite{bc2s2}
\\
$3~^1S_0$ &$0^-$&7228&7250&7226&7254&7239&&
\\
$3~^3S_1$&$1^-$ &7272&7272&7258&7280&7252&&
\\
$1~^3P_0$ &$0^+$&6705~({\scriptsize{input}})&6706&6699&6693&6714&6712&
\\
$1~P_1$ &$1^+$&6739~({\scriptsize{input}})&6741&6743&6731&6757&6736&
\\
$1~P'_1$ &$1^+$&6748&6750&6750&6739&6776&&
\\
$\theta_{1P}$&&$-57.8^\circ$($32.2^\circ$)&$22.4^\circ$&$20.5^\circ$&$18.7^\circ$&$35.5^\circ$&$33.4\pm1.5^\circ$ \cite{davies}&
\\
$1~^3P_2$ &$2^+$&6762~({\scriptsize{input}})&6768&6761&6751&6787&&
\\
$2~^3P_0$ &$0^+$&7112&7122&7094&7105&7107&&
\\
$2~P_1$ &$1^+$&7144&7145&7134&7136&7134&&
\\
$2~P'_1$&$1^+$ &7149&7150&7147&7144&7150&&
\\
$\theta_{2P}$&&$-59.1^\circ$($30.9^\circ$)&$18.9^\circ$&$23.2^\circ$&$21.2^\circ$&$38.0^\circ$&&
\\
$2~^3P_2$ &$2^+$&7163&7164&7157&7155&7160&&
\\
$3~^3P_0$&$0^+$ &7408&&7474&7437&7420&&
\\
$3~P_1$ &$1^+$&7440&&7500&7465&7441&&
\\
$3~P'_1$ &$1^+$&7442&&7510&7474&7458&&
\\
$\theta_{3P}$&&$-60.1^\circ$($29.9^\circ$)&&&&$39.7^\circ$&&
\\
$3~^3P_2$ &$2^+$&7456&&7524&7483&7464&&
\\
\hline\hline
\end{tabular*}
\end{table*}

\begin{table*}[hbt]
Continued: Mass spectra and mixing angles of the $B_c$ system.
\begin{tabular*}{\textwidth}{@{}c@{\extracolsep{\fill}}ccccccc}
\hline \hline
$n~^{2S+1}L_J$&$J^P$&\rm{ours}&\cite{godfrey}&\cite{Ebert}&\cite{eichten}&\cite{zhong}\\ \hline

$1~{}^3D_1$ &$1^-$&7014~({\scriptsize{$S-P-D$}})&7028&7021&7007&7020
\\
$2~{^3}D_1$ &$1^-$&7335~({\scriptsize{$S-P-D$}})&&7392&7347&7336
\\
%$3~{}^3D_1$&$1^-$ &7586~($S-D$)&&7732&&7611&&
%\\
$1~{}^3F_2$&$2^+$ &7239~({\scriptsize{$P-D-F$}})&7269&7232&7234&7235
\\
$2~{}^3F_2$ &$2^+$&7508~({\scriptsize{$P-D-F$}})&&7618&&7518
\\
%$3~{}^3F_2$ &$2^+$&7735~($P-F$)&&&&7730&&
%\\
$1~^3D_3$ &$3^-$&7035~({\scriptsize{input}})&7045&7029&7011&7030
\\
$1~D_2$ &$2^-$&7025~({\scriptsize{input}})&7036&7025&7006&7024
\\
$1~D'_2$ &$2^-$&7029&7041&7026&7016&7032
\\
$\theta_{1D}$&&$-58.2^\circ$($31.8^\circ$)&$44.5^\circ$&$-35.9^\circ$&$-49.2^\circ$&$45.0^\circ$
\\
$2~^3D_3$ &$3^-$&7355&&7405&7351&7348
\\
$2~D_2$ &$2^-$&7345&&7399&7339&7343
\\
$2~D'_2$ &$2^-$&7349&&7400&7359&7347
\\
$\theta_{2D}$&&$-57.4^\circ$($32.6^\circ$)&&&$-40.3^\circ$&$45.0^\circ$
\\
\hline\hline
\end{tabular*}
\end{table*}

\subsection{Partial waves in the $0^-$ and $0^+$ states}
\subsubsection{$0^-$ state}
\begin{figure}
\centering
\includegraphics[width=0.43\textwidth]{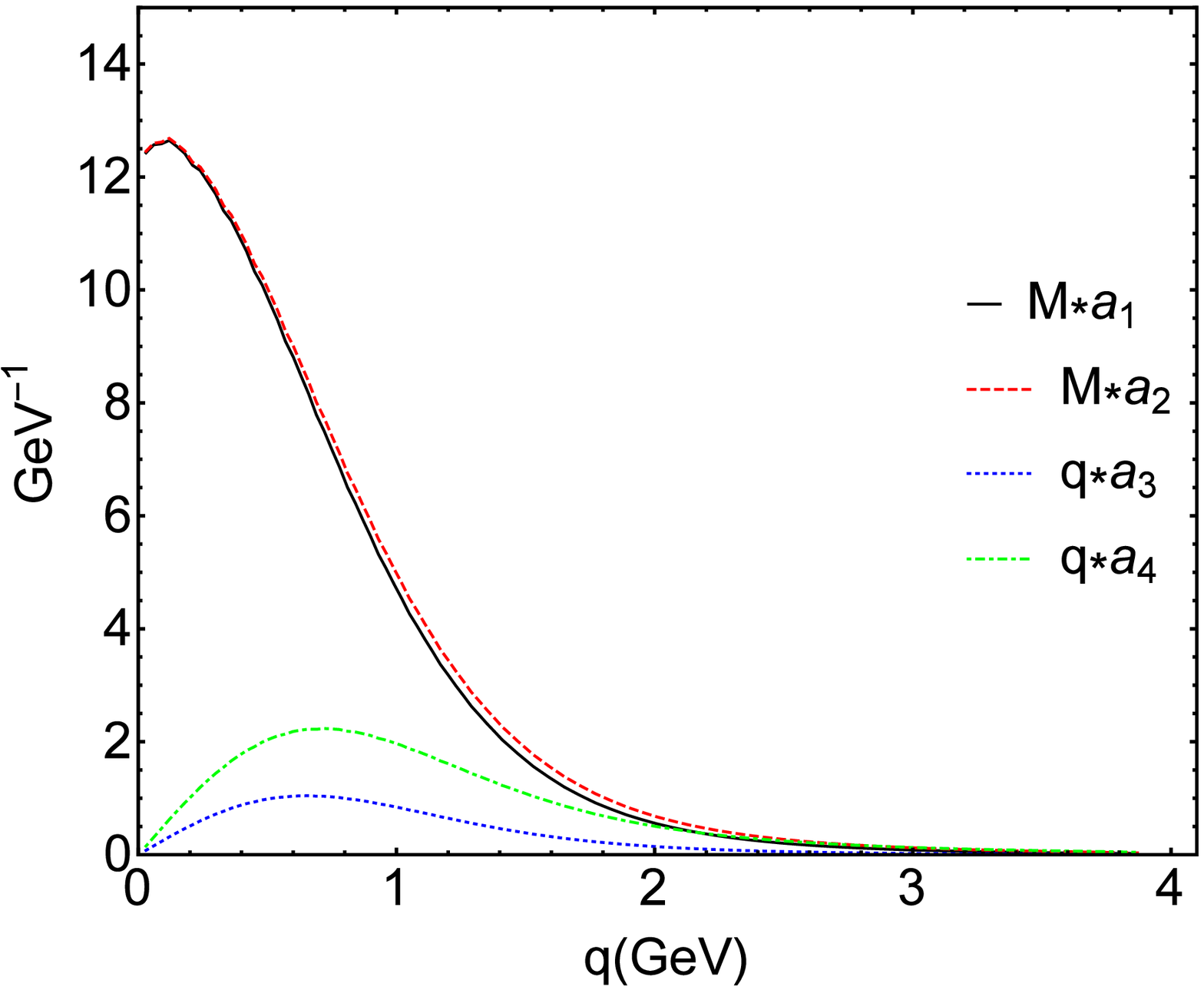}
\includegraphics[width=0.43\textwidth]{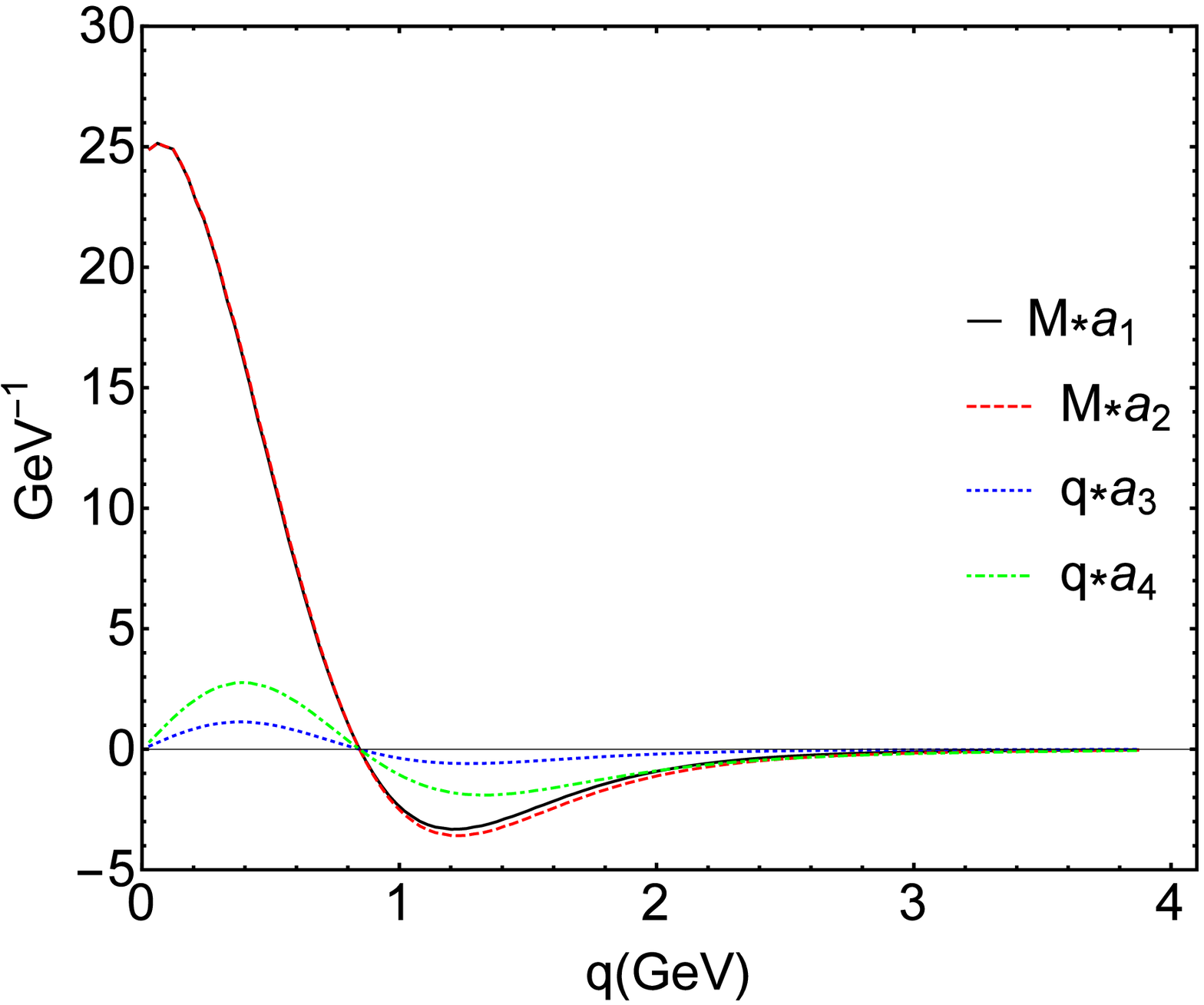}
\caption{The $0^-$ wave functions of the ground state $B_c(1S)$ (left) and the first excited state $B_c(2S)$ (right). $a_1$ and $a_2$ terms are $S$ waves; $a_3$ and $a_4$ terms are $P$ waves.} \label{f0-}
\end{figure}

Using the numerical result of the normalization formula Eq.(\ref{0-nor}) and the left side of Eq.(\ref{0-pure}), we get the proportions of the $S$ and $P$ partial waves in the $0^-$ $B_c$ system. Their ratios are $S:P=1:0.082$, $1:0.091$, $1:0.097$ and $1:0.11$ for $1S$, $2S$, $3S$, and $4S$ states, respectively. So the $0^-$ $B_c(nS)$  is not a pure $n^1S_0$ state, but contains a small amount of $P$ wave component.
{In Figure \ref{f0-}, we plot the wave functions of the ground state $B_c(1S)$ and the first radial excited state $B_c(2S)$. We can see clearly that, in addition to the dominant $S$ waves $a_1$ and $a_2$, there are small $P$ wave $a_3$ and $a_4$ terms.}

\subsubsection{$0^+$ state}
\begin{figure}
\centering
\includegraphics[width=0.43\textwidth]{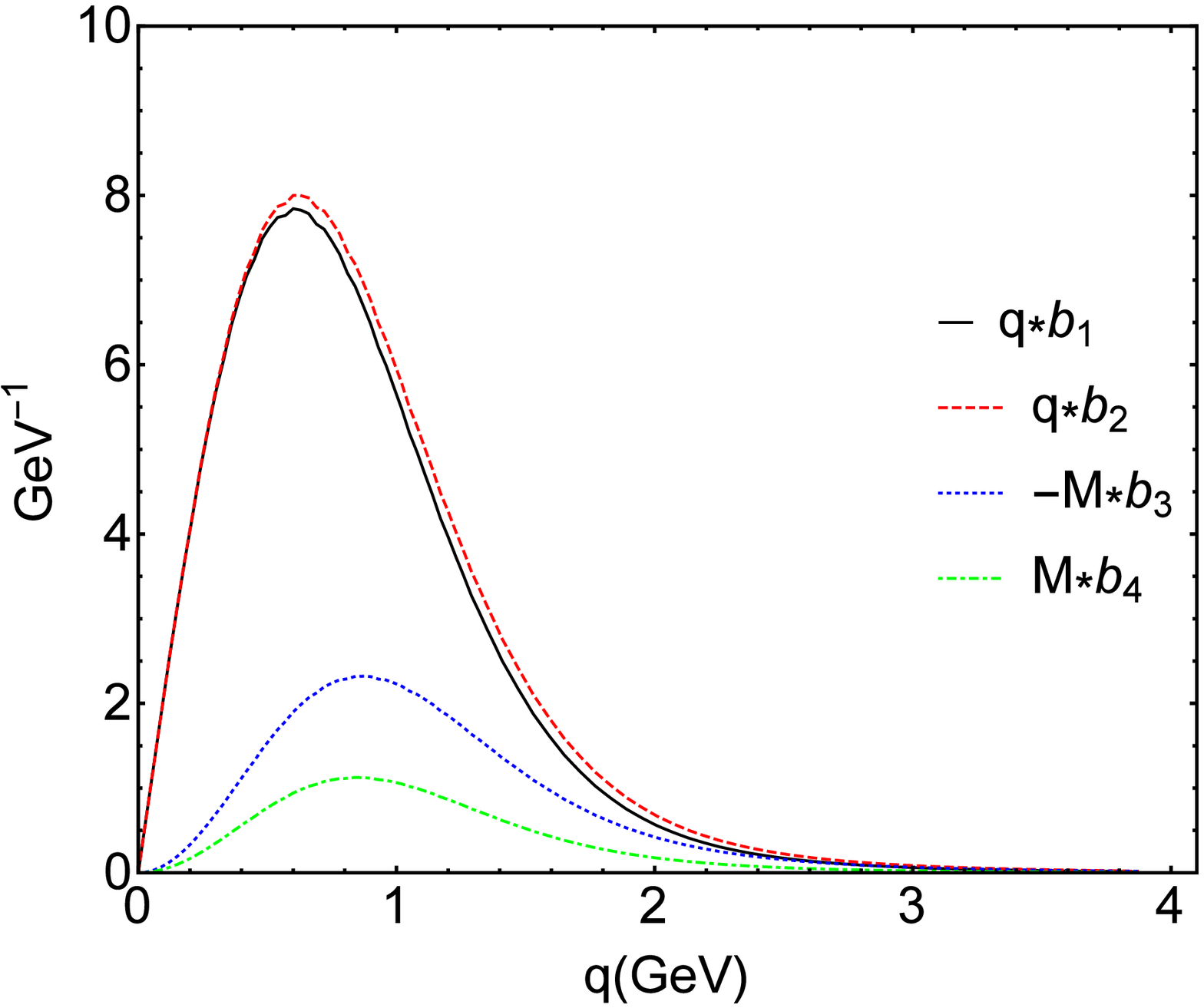}
\includegraphics[width=0.43\textwidth]{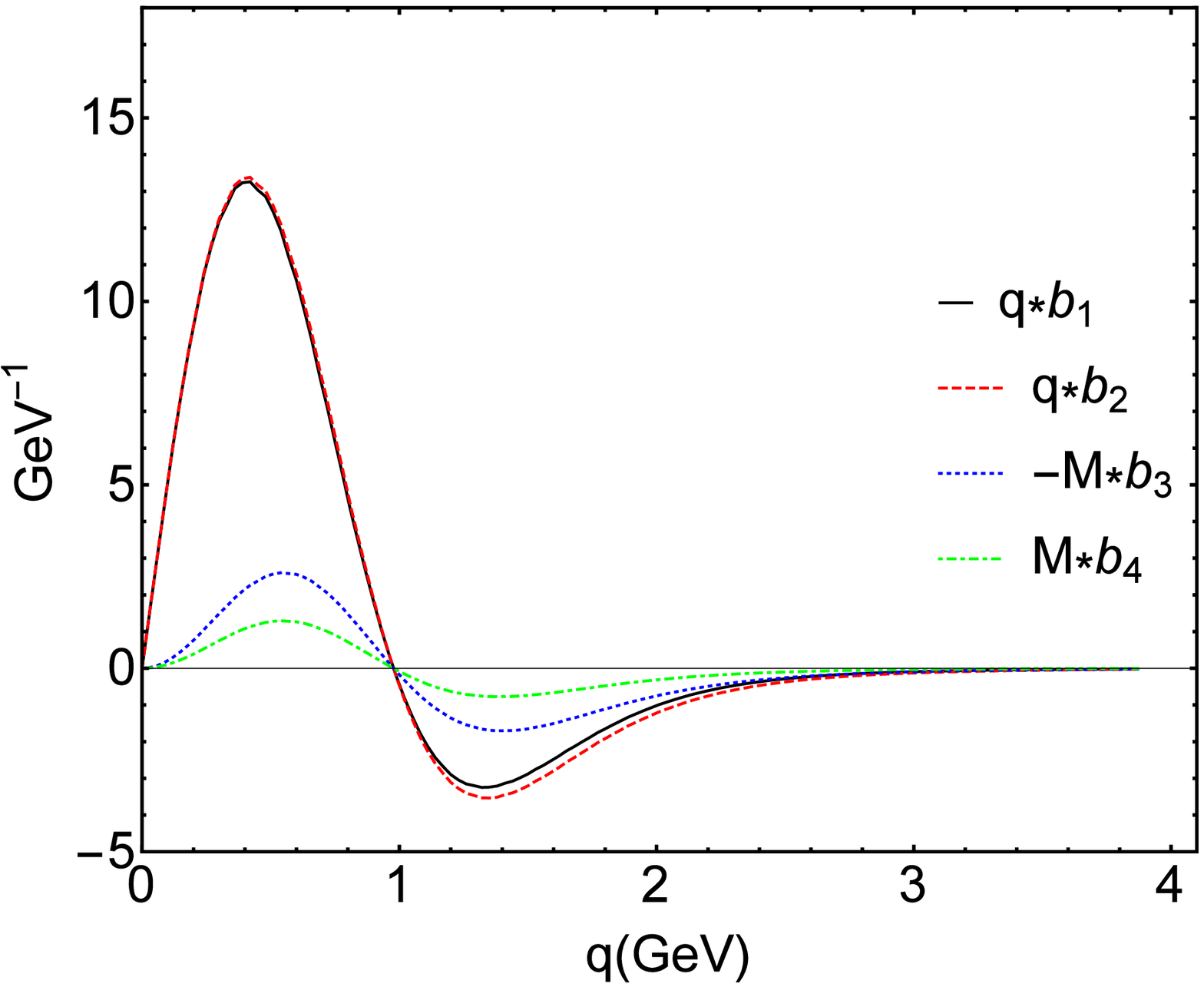}
\caption{The $0^+$ wave functions of the state $B^*_{c0}(1P)$ (left) and its radial excited state $B^*_{c0}(2P)$ (right). $b_1$ and $b_2$ terms are $P$ waves; $b_3$ and $b_4$ terms are $S$ waves.} \label{f0+}
\end{figure}

Similar to the case of $0^-$ state, using the normalization formula Eq.(\ref{0+nor}) and the left side of Eq.(\ref{0+pure}), we obtain the ratio $P:S=1:0.097$, $1:0.10$, $1:0.11$ and $1:0.12$ for $1P$, $2P$, $3P$, and $4P$ $0^+$ $B^*_{c0}$ states, respectively. Then we conclude that the $0^+$ $B^*_{c0}(nP)$ meson is a $n^3P_0$ wave dominant state, and its wave function also has a small amount of $S$ wave. {The corresponding wave functions of the $B^*_{c0}(1P)$ and $B^*_{c0}(2P)$ are shown in Fig. \ref{f0+}, where the dominant terms $b_1$ and $b_2$ are $P$ waves, small terms $b_3$ and $b_4$ are $S$ waves.}

\subsection{Partial waves in the $1^-$, $2^+$ and $3^-$ states}
\subsubsection{$1^-$ state}
The solutions of $1^-$ $B^*_c$ state divide into two categories. The first category of the wave functions includes the first solution, second solution, fourth solution, etc. We obtain the following ratios $S:P:D=1:0.09:0.037$, $1:0.097:0.044$ and $1:0.10:0.017$, which means they are $1S$, $2S$ and $3S$ dominant states, respectively. The other two kinds of partial waves, especially the $D$ wave component,  are very small,  which can be safely ignored.

The second category includes the third and fifth solutions. Their ratios are $S:P:D= -0.576:0.48:1$ and $-0.575:0.48:1$, which indicate that they are $D$ wave dominant states, but have sizable $S$ wave and $P$ wave components. Usually they are considered as $S-D$ mixing states in a non-relativistic model, but our results show that they also include a large amount of $P$ wave component, so in Table \ref{mass}, we mark them as the $S-P-D$ mixing states.

\begin{figure}
\centering
\includegraphics[width=0.32\textwidth]{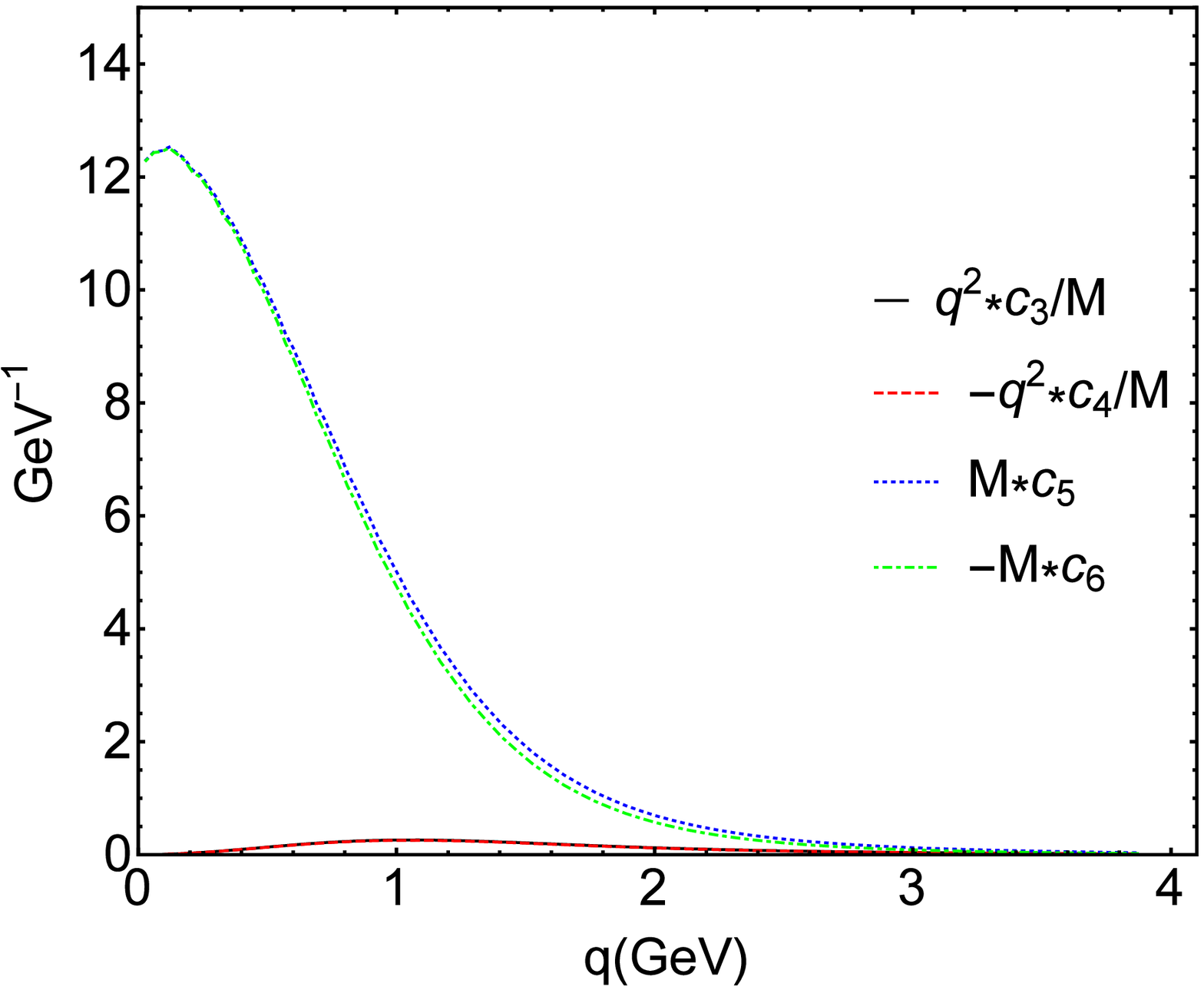}
\includegraphics[width=0.32\textwidth]{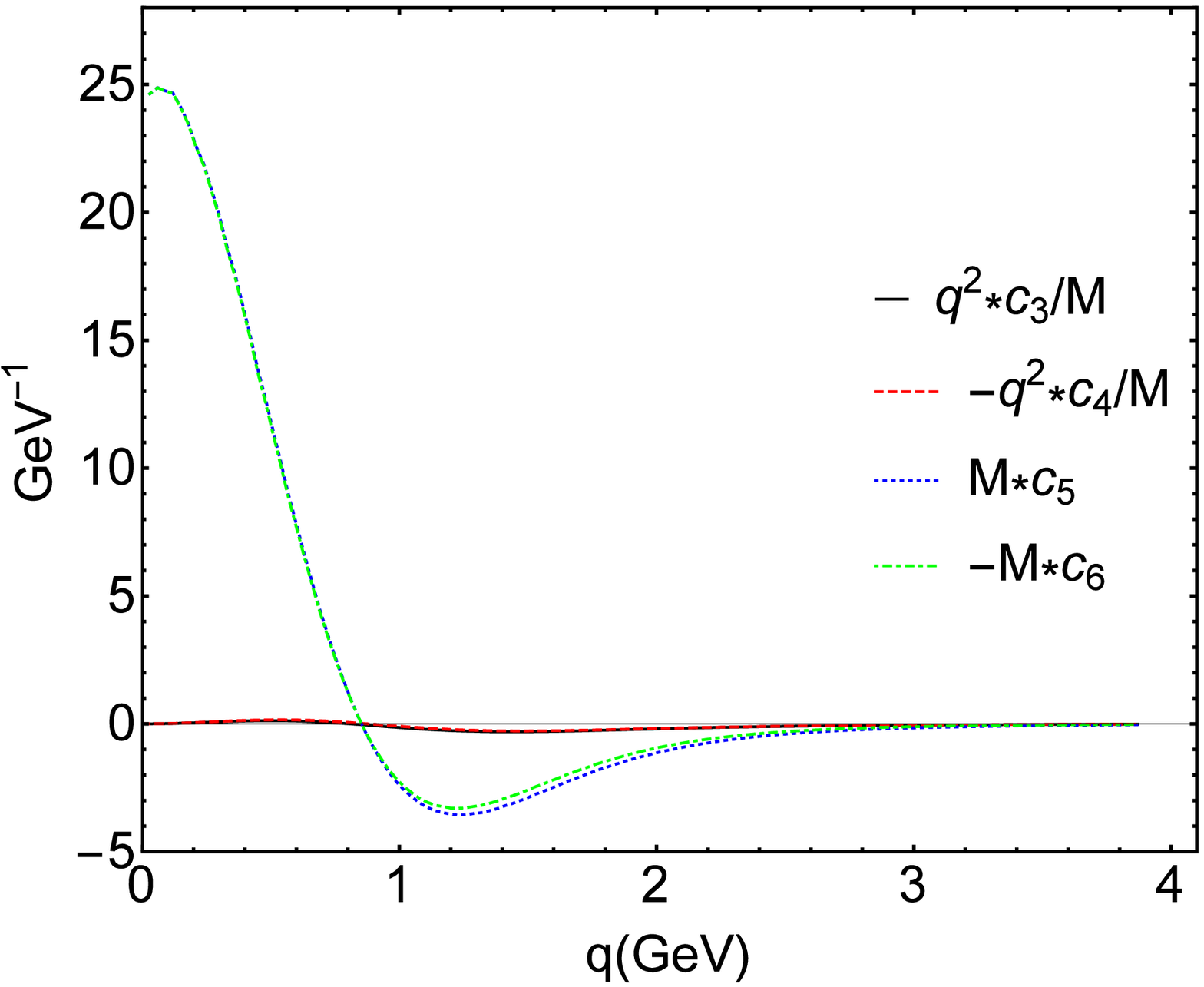}
\includegraphics[width=0.32\textwidth]{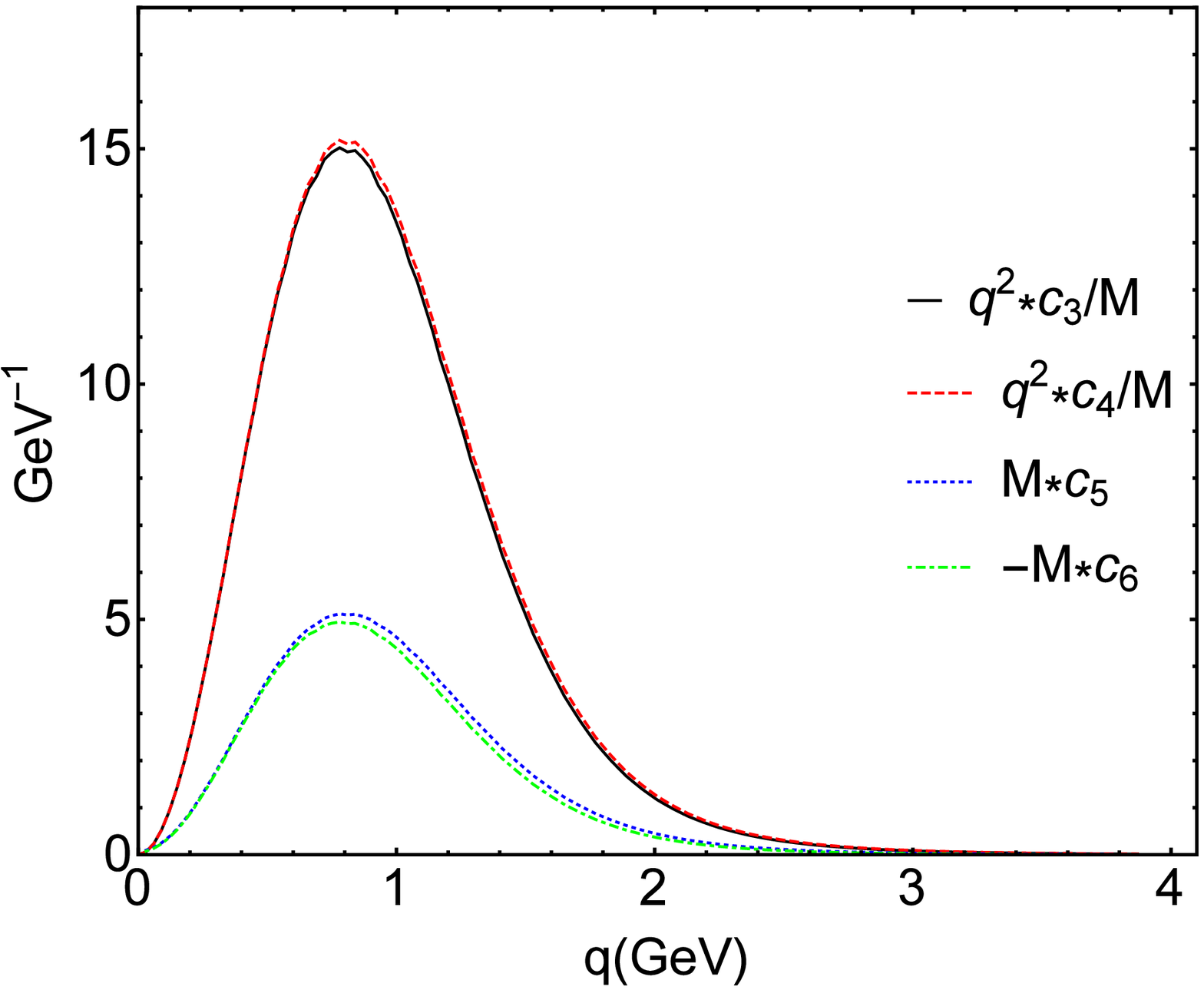}
\caption{The $1^-$ wave functions of the $B^*_c(1S)$ (left), $B^*_c(2S)$ (middle) and the $S-P-D$ mixture $B^*_c(1D)$ (right). $c_3$ and $c_4$ terms are $D$ waves; $c_5$ and $c_6$ terms are $S$ waves.} \label{f1-}
\end{figure}

{In Figure \ref{f1-}, we plot the wave functions of the ground state $B^*_c(1S)$, the first radial excited state $B^*_c(2S)$, and the $S-P-D$ mixing state. Since the last one is the $1^3D_1$ dominant state, we label it as $B^*_c(1D)$. Unlike the $0^-$ and $0^+$ cases, we do not give all the partial waves in Fig. \ref{f1-}, but only show four independent wave functions, which are the $S$ and $D$ waves, and the $P$ waves which are not shown are their functions.}

\subsubsection{$2^+$ state}
For the wave function of $2^+$ $B^*_{c2}$ system, we obtain the ratios $P:D:F=1:0.10:0.039$, $1:0.11:0.049$ and $1:0.13:0.055$ for the first, second and forth solutions, respectively. So they are $1^3P_2$, $2^3P_2$, $3^3P_2$ dominant state $B^*_{c2}(nP)$, and have small amounts of $D$ and $F$ wave components.

\begin{figure}
\centering
\includegraphics[width=0.32\textwidth]{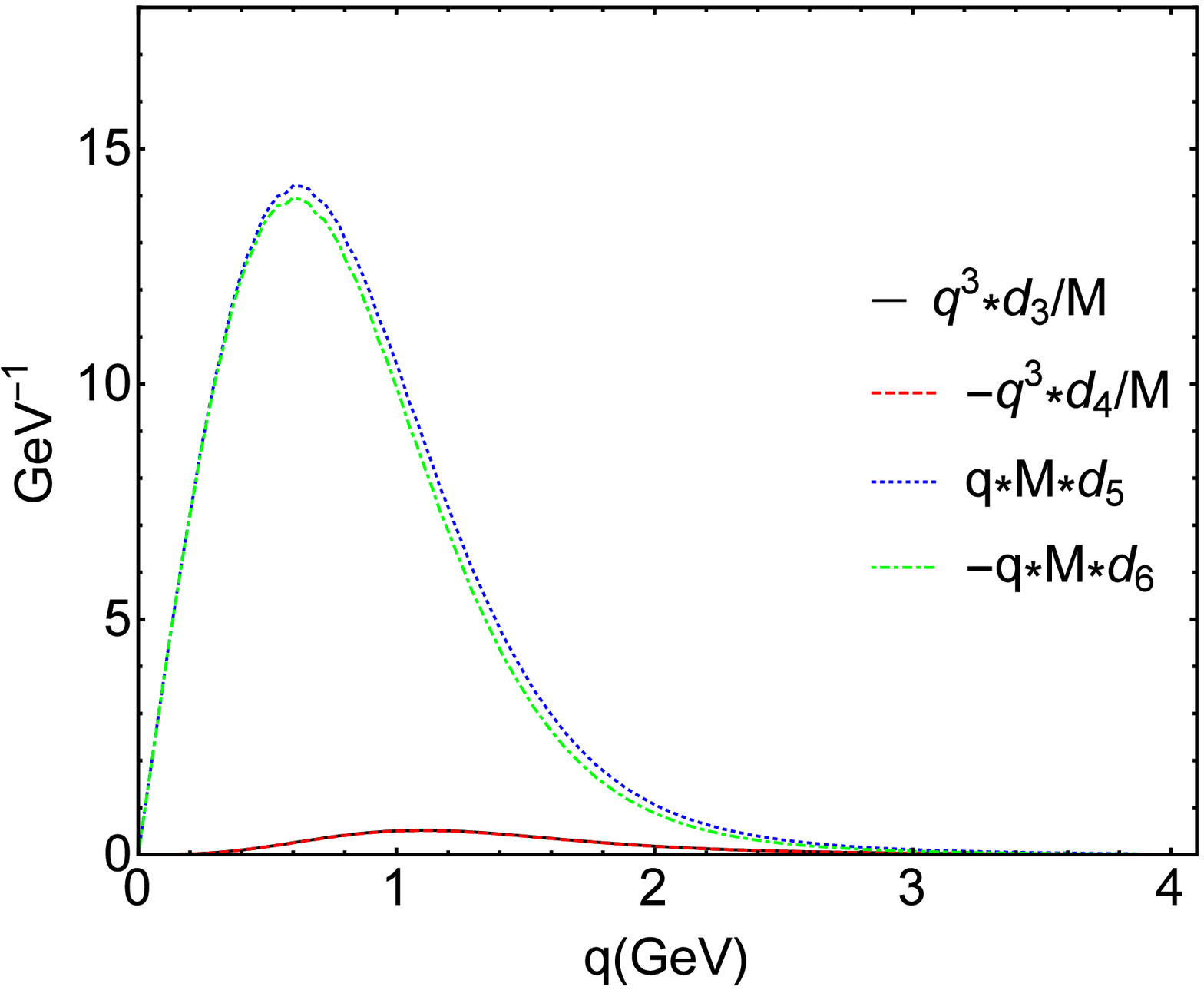}
\includegraphics[width=0.32\textwidth]{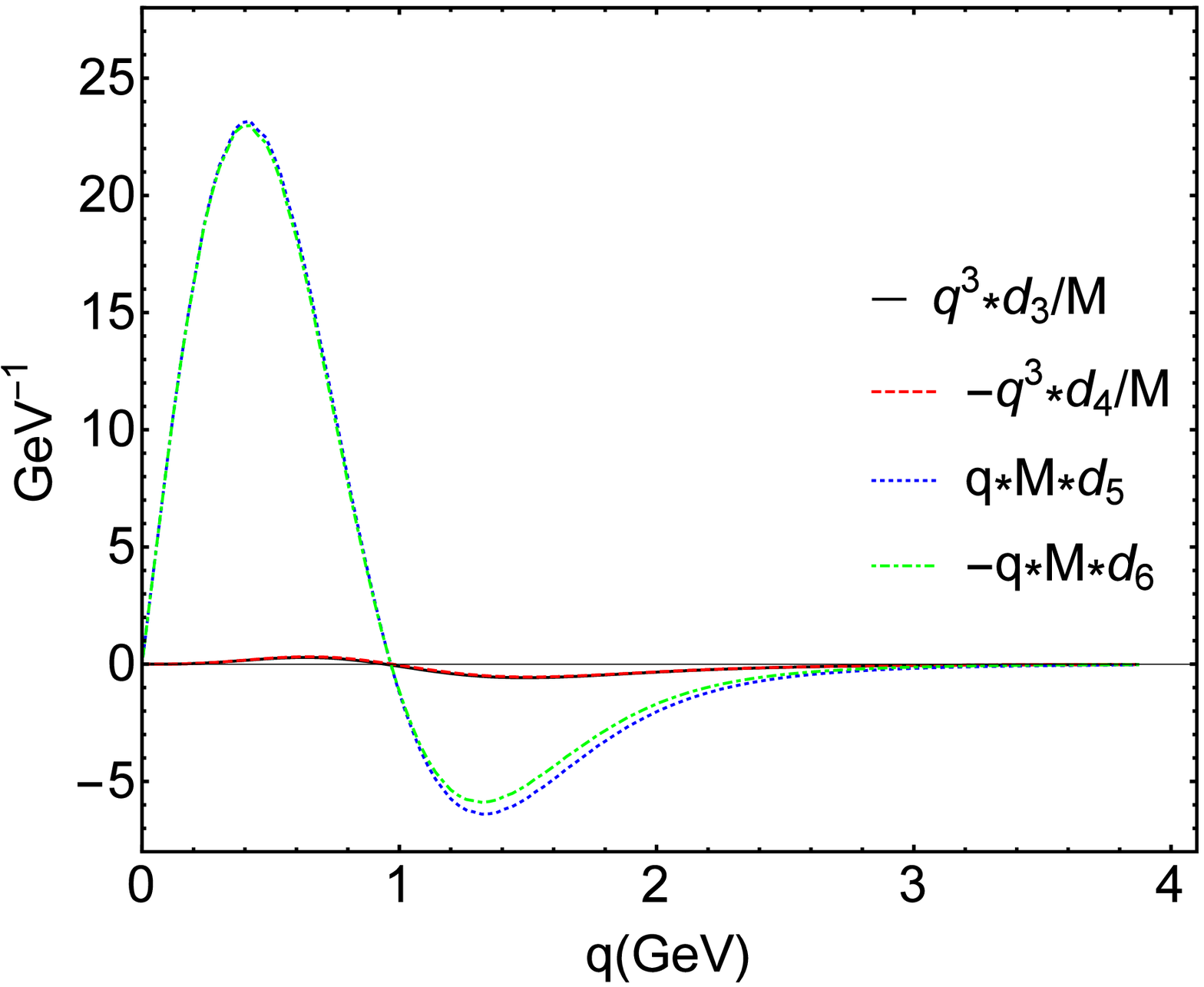}
\includegraphics[width=0.32\textwidth]{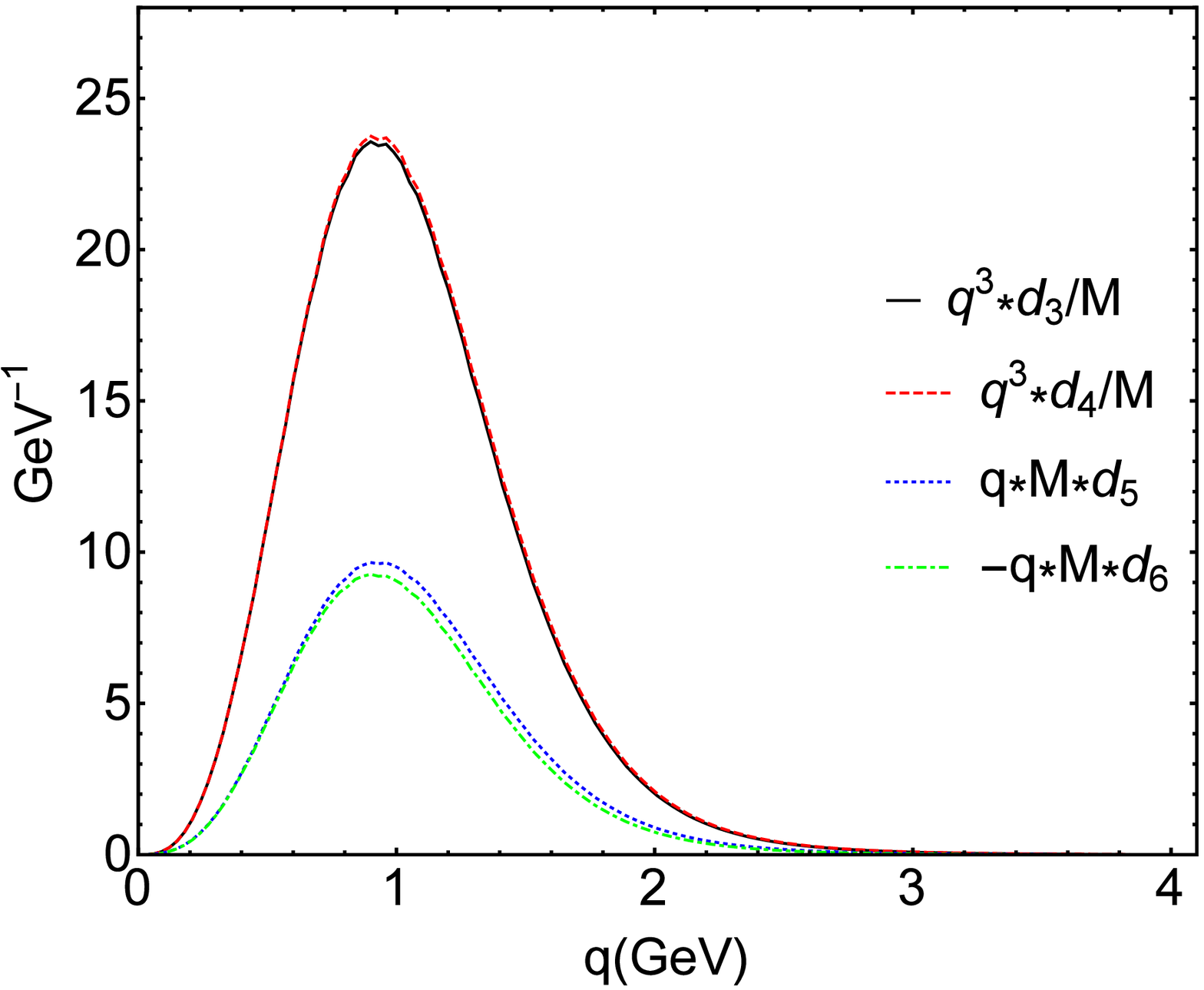}
\caption{The $2^+$ wave functions of the $B^*_{c2}(1P)$ (left), $B^*_{c2}(2P)$ (middle) and the $P-D-F$ mixture $B^*_{c2}(1F)$ (right). $d_3$ and $d_4$ terms are $F$ waves; $d_5$ and $d_6$ terms are $P$ waves.} \label{f2+}
\end{figure}

The third and fifth solutions have the ratios $P:D:F=-0.633 : 0.50 : 1$ and $-0.635 : 0.50 : 1$, so they are $n^3F_2$ wave dominant states, but contain large amounts of $P$-wave and $D$-wave components. Usually they are believed to be the $P-F$ mixing states, but our results show that they also include a sizable amount of $D$-wave component, so in Table \ref{mass}, we mark them as the $P-D-F$ mixing states. {While in Figure \ref{f2+},  where the wave functions for the first three solutions are drawn, the $^3F_2$ wave dominant mixing state is labeled as $B^*_{c2}(1F)$.}

\subsubsection{$3^-$ state}
Similarly, for the first, second, and forth solutions of the $3^-$ $B^*_{c3}$  system, we obtain the ratios $D:F:G=1:0.11:0.043$, $1:0.13:0.052$ and $1:0.14:0.058$, respectively, which means they are respectively $1D$, $2D$ and $3D$ dominant states $B^*_{c3}(nD)$ with a small amount of $F$ and $G$ waves components.

\begin{figure}
\centering
\includegraphics[width=0.32\textwidth]{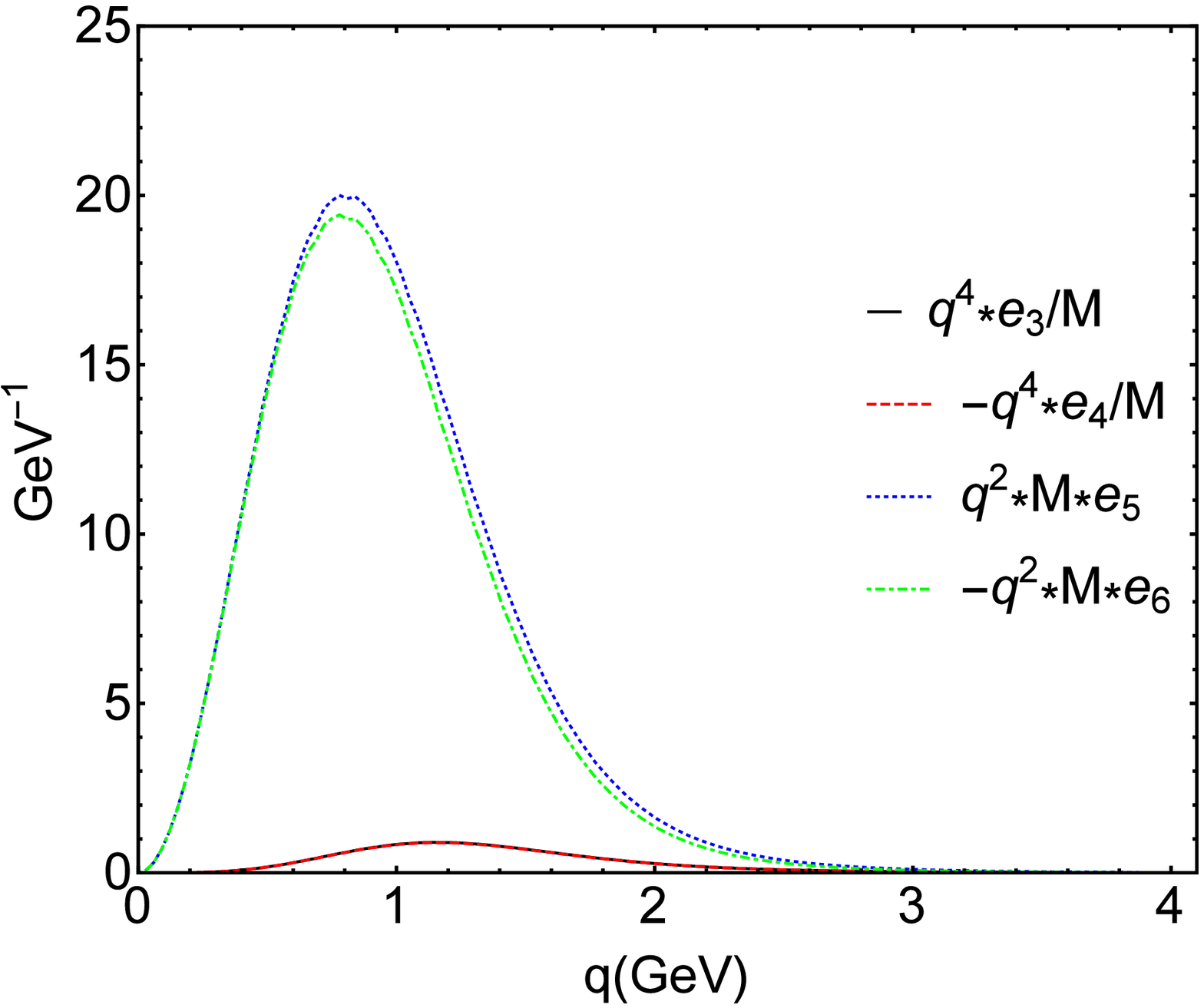}
\includegraphics[width=0.32\textwidth]{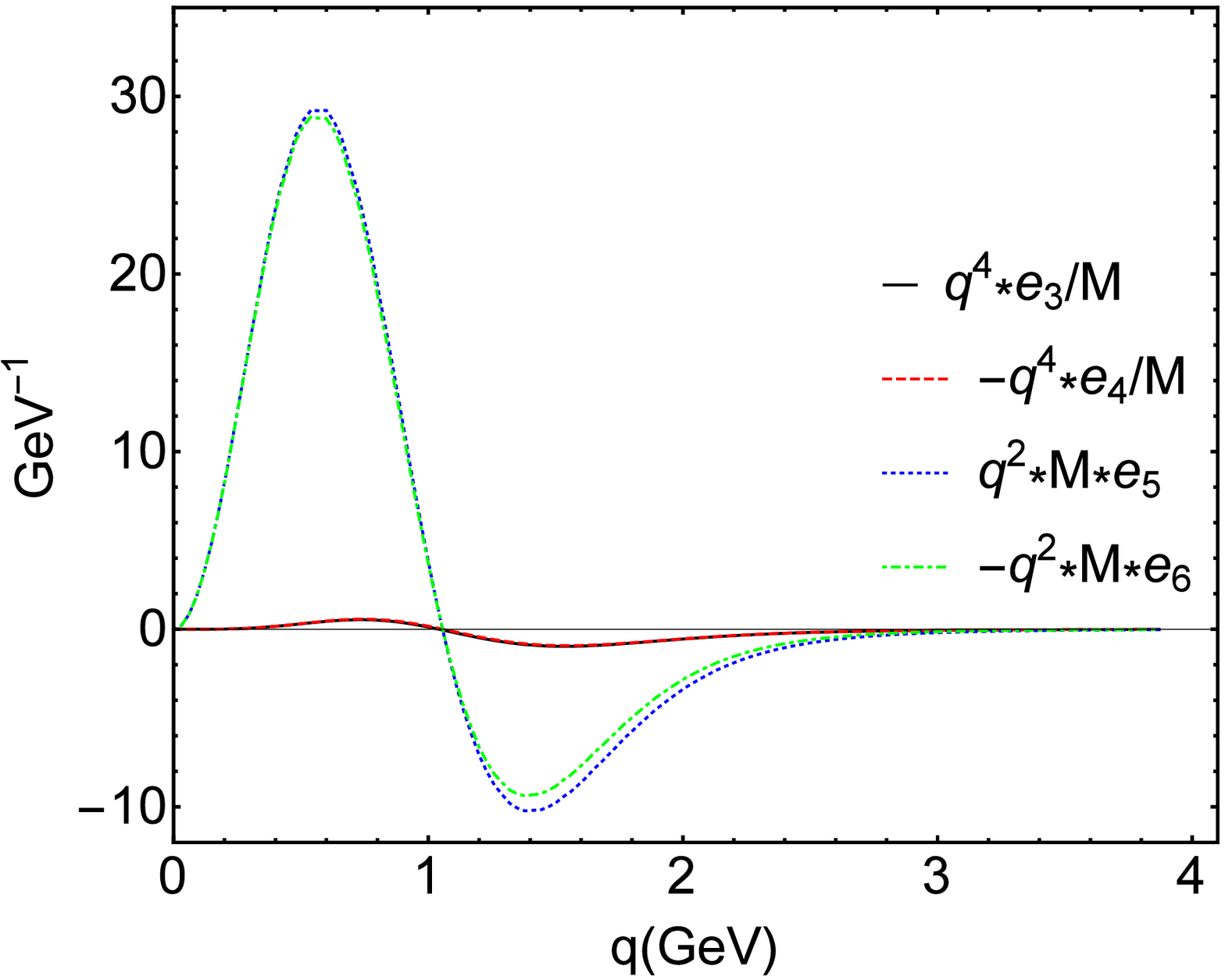}
\includegraphics[width=0.32\textwidth]{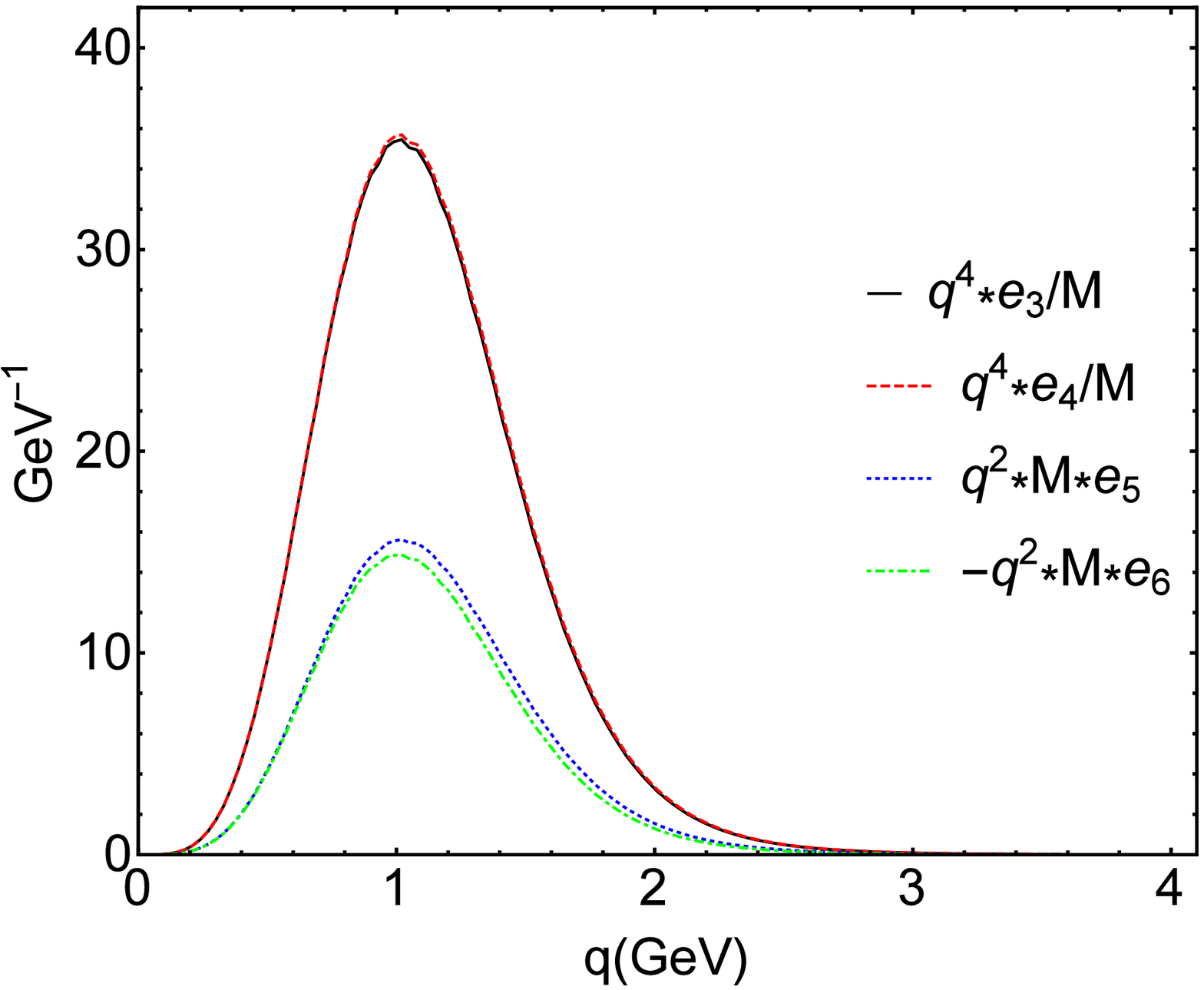}
\caption{The $3^-$ wave functions of the $B^*_{c3}(1D)$ (left), $B^*_{c3}(2D)$ (middle) and the $D-F-G$ mixture $B^*_{c3}(1G)$ (right). $e_3$ and $e_4$ terms are $G$ waves; $e_5$ and $e_6$ terms are $D$ waves.} \label{f3-}
\end{figure}

The third and fifth solutions have the ratios $D:F:G=-0.654:0.505:1$ and $ -0.654:0.508:1$. One can see that the $G$ wave is dominant, but the other two components are also important. So they are $D-F-G$ mixing states, and labeled as the state $B^*_{c3}(nF)$ in Figure \ref{f3-}.

\subsection{Partial waves in the $1^+$ and $2^-$ states}

\subsubsection{Partial waves and the ${}^1P_1-{}^3P_1$ mixing angle of the $1^+$ state}

\begin{figure}
\centering
\includegraphics[width=0.43\textwidth]{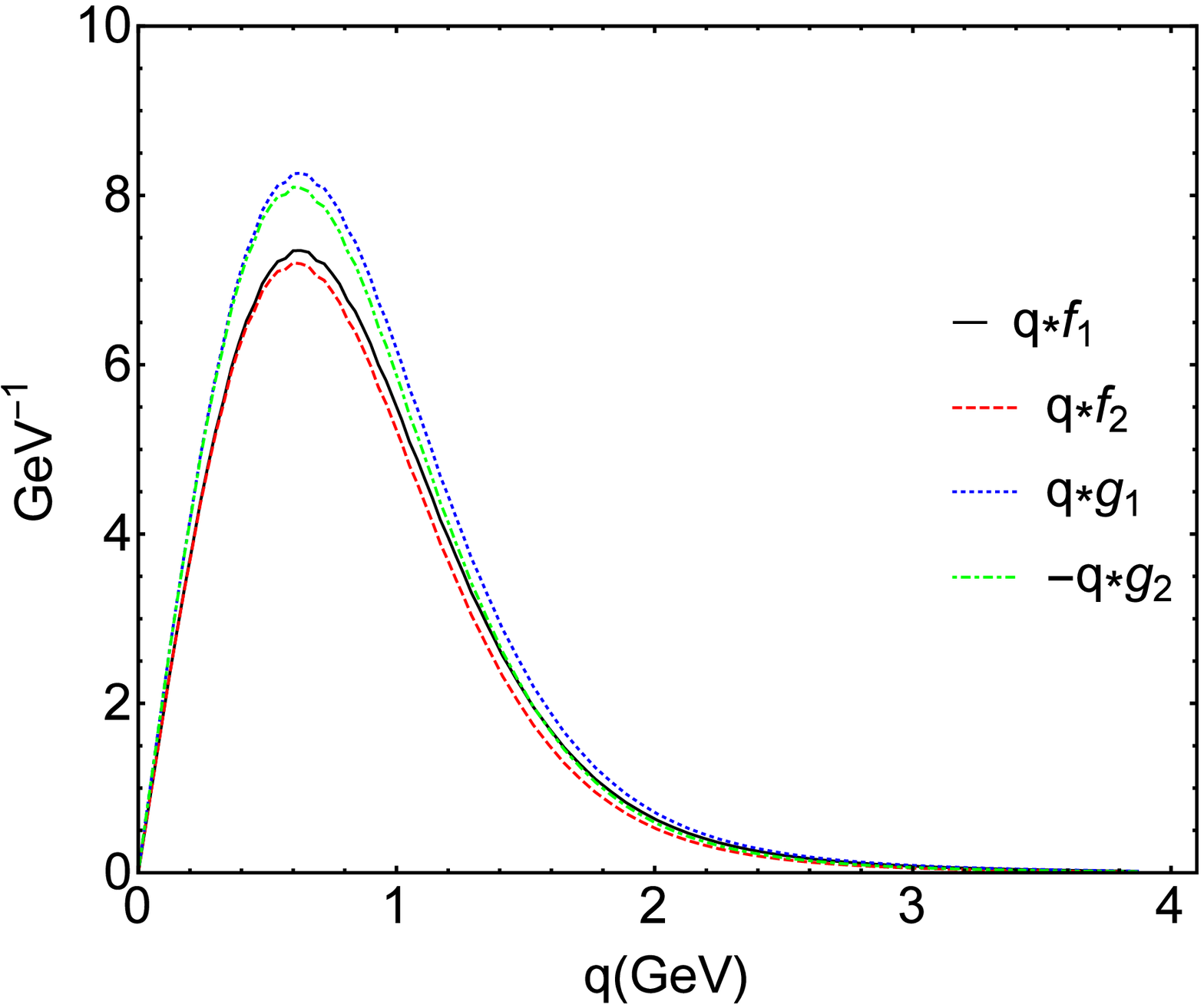}
\includegraphics[width=0.43\textwidth]{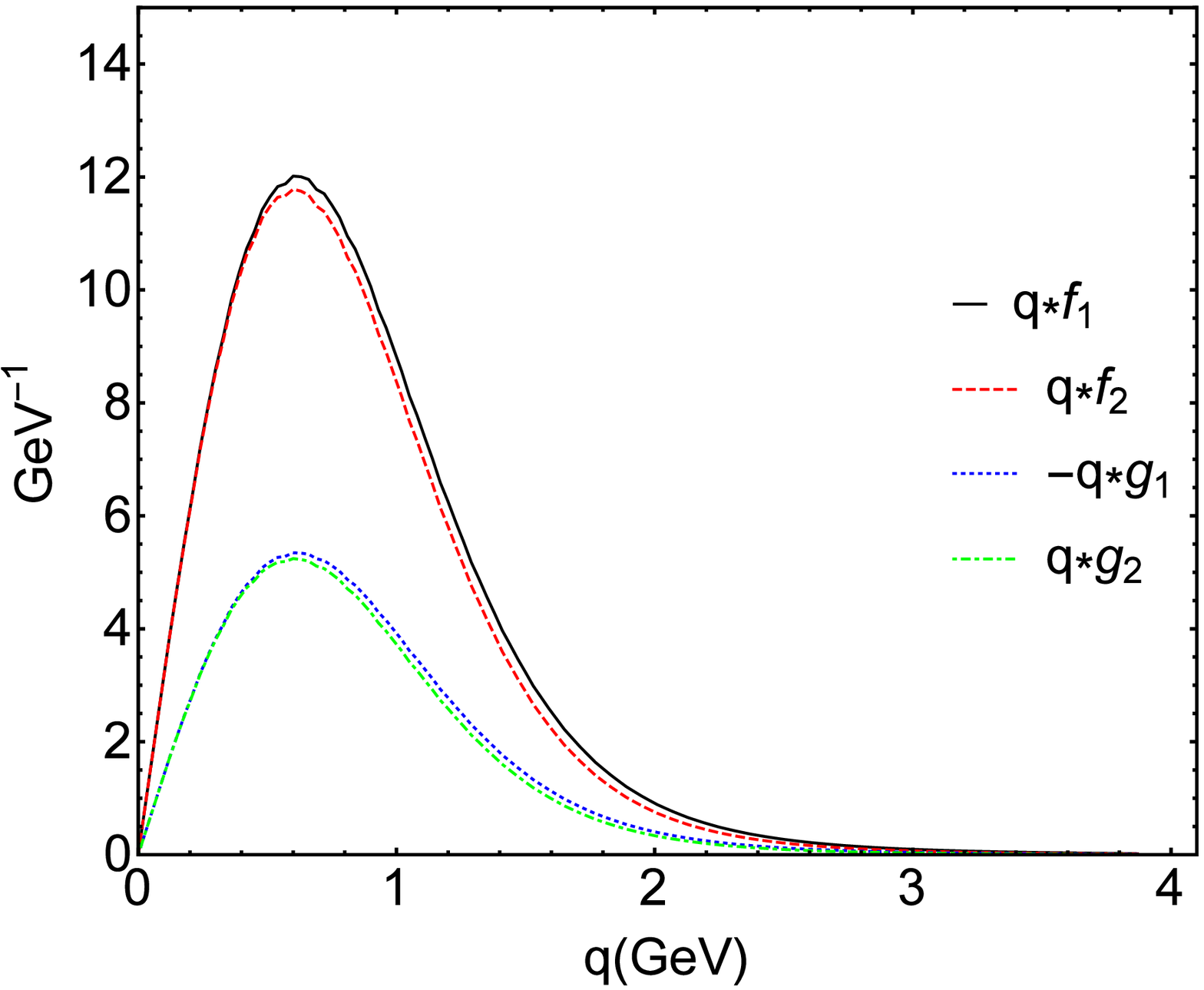}
\caption{The $1^+$ wave functions of the $1^1P_1-1^3P_1$ mixing states $B_{c1}(1P)$ (left) and $B'_{c1}(1P)$ (right). $f_1$ and $f_2$ terms are $^1P_1$ waves; $g_1$ and $g_2$ terms are $^3P_1$ waves.} \label{f1+1P}
\end{figure}

\begin{figure}
\centering
\includegraphics[width=0.43\textwidth]{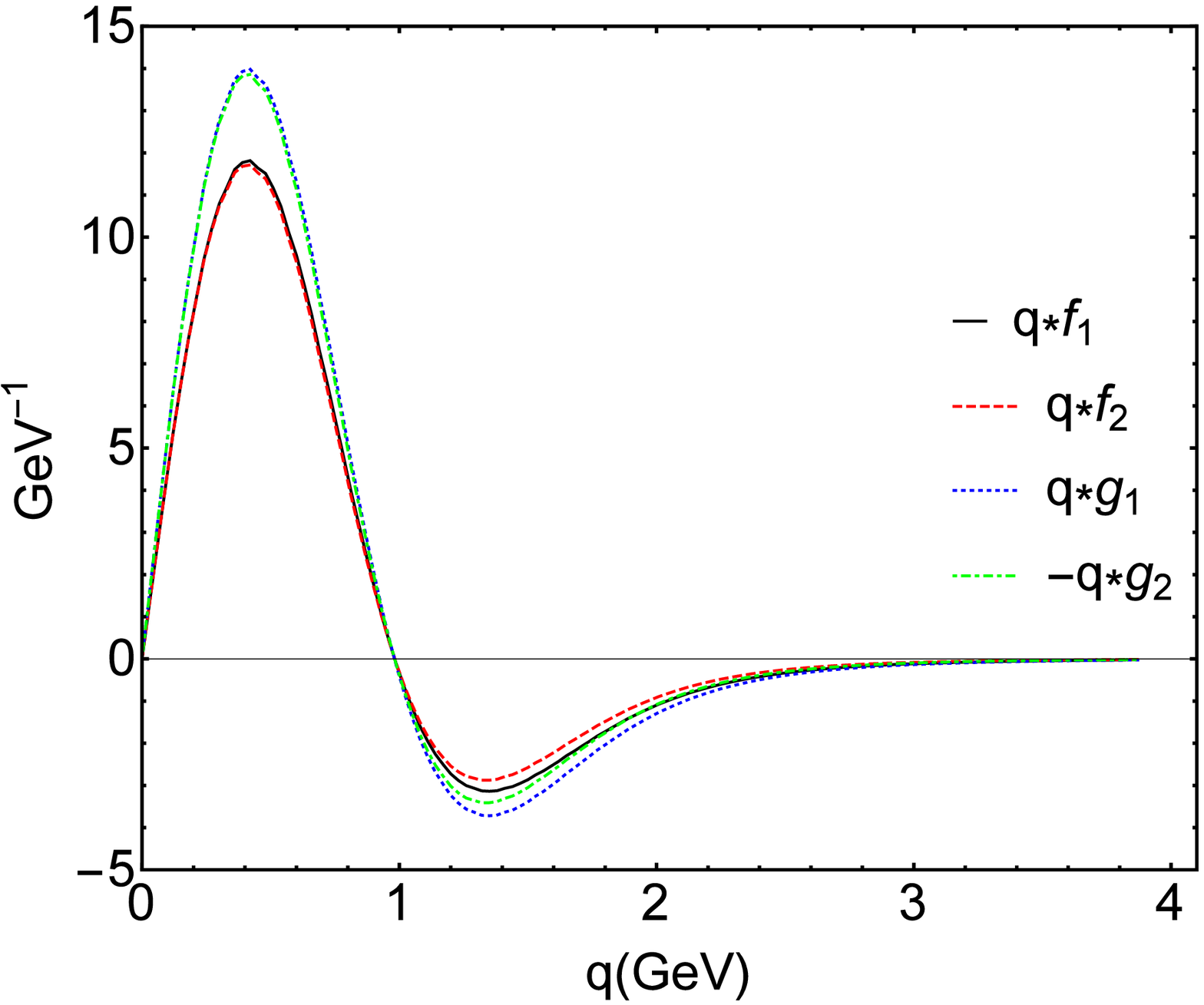}
\includegraphics[width=0.43\textwidth]{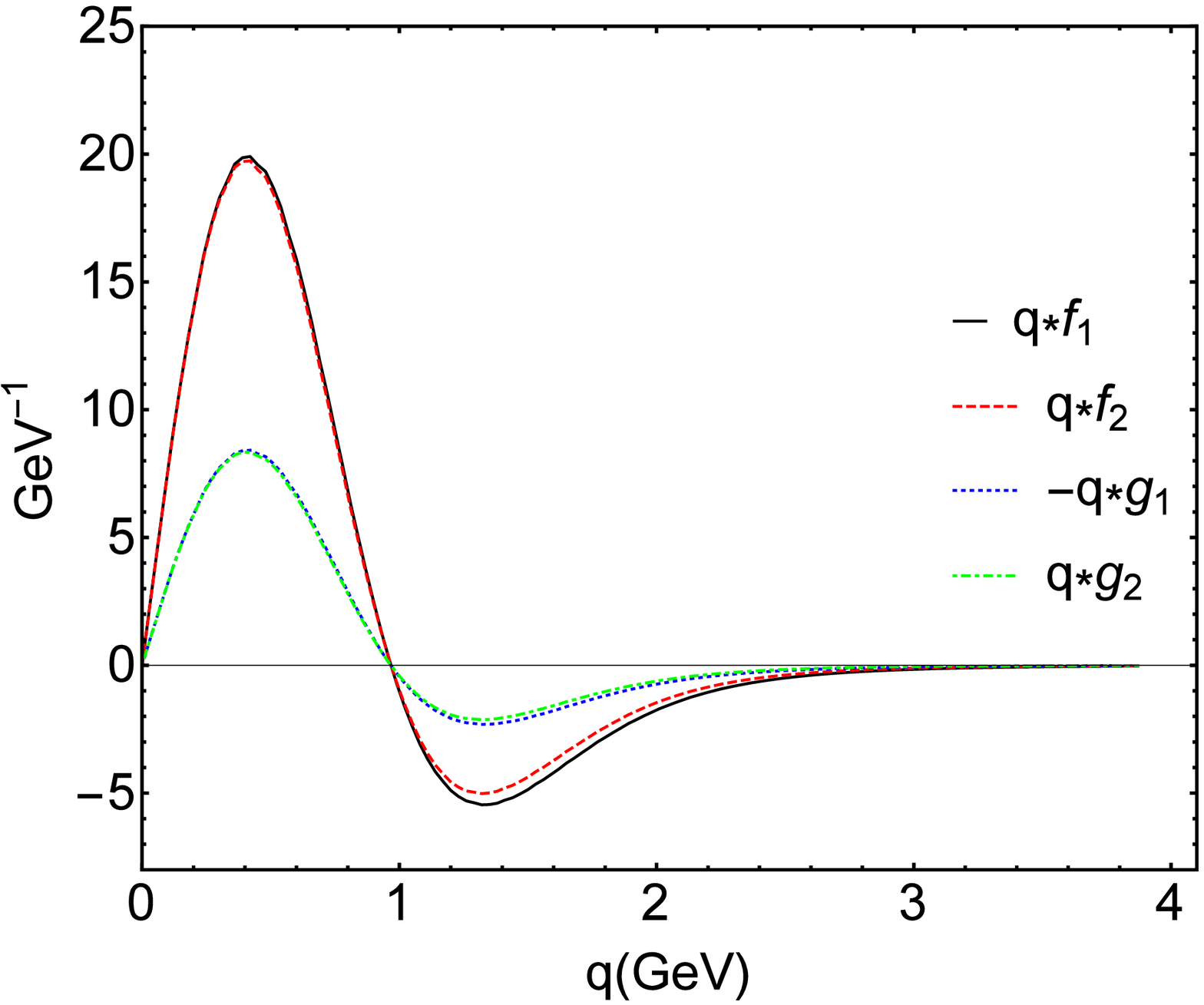}
\caption{The $1^+$ wave functions of the $2^1P_1-2^3P_1$ mixing states $B_{c1}(2P)$ (left) and $B'_{c1}(2P)$ (right). $f_1$ and $f_2$ terms are $^1P_1$ waves; $g_1$ and $g_2$ terms are $^3P_1$ waves.} \label{f1+2P}
\end{figure}

Different from the case of the previous states, the solutions of the $1^+$ state equation appear in pairs. For example, the first and second solutions are all $1P$ states. Using the numerical results of the normalization condition Eq.(\ref{1+nor}), we obtain the ratios $1{}^1P_1 : 1{}^3P_1=0.284:0.716$ and $0.716:0.284$, and the mixing angle between $1{}^1P_1$ and $1{}^3P_1$ being  $\theta_{1P}=-57.8^\circ$ or $32.2^\circ$. The third and forth solutions are all $2P$ states, for which we obtain $2{}^1P_1 : 2{}^3P_1=0.263:0.737$ and $0.737:0.263$,  $\theta_{2P}=-59.1^\circ$ or $30.9^\circ$. The fifth and sixth solutions are $3P$ states, with $3{}^1P_1 :3 {}^3P_1=0.248:0.752$ and $0.752:0.248$, $\theta_{3P}=-60.1^\circ$ or $29.9^\circ$. {In Figures \ref{f1+1P} and \ref{f1+2P}, we show the wave functions of $B^{(\prime)}_{c1}(1P)$ and their excited states $B^{(\prime)}_{c1}(2P)$, respectively.}

The values of the mixing angles are also listed in Table \ref{mass}. Our result $\theta_{1P}=32.2\wang{^\circ}$ is in good agreement with the Lattice result $\theta_{1P}=33.4\pm1.5^\circ$ \cite{davies} and the constituent quark model result $\theta_{1P}=35.5^\circ$ \cite{zhong}. It can be seen that the results of most theoretical models are quite different. This is because when calculating the mixing angle, such methods use the interaction potential, whose complete form is difficult to give. But we use the wave function to make the calculation, which is relatively complete.

We point out that, the states considered here are not pure $P$ waves, but mixed with small amount of $D$ wave components. The $f_1$, $f_2$, $g_1$ and $g_2$ terms in Eq.(\ref{1+wave}) are $P$ waves, but $f_3$, $f_4$, $g_3$ and $g_4$ terms are actually $D$ waves. The ratio for the first and second solutions, which are $1P$ dominant states with a small amount of $D$ wave, is $P : D=1:0.0971$. Similarly, we get
$P : D=1:0.0936$ for two $2P$ dominant states, and $P : D=1:0.104$ for two $3P$ dominant states. If we ignore the small $D$ wave contribution, and only consider the dominant $P$ wave, the corresponding mixing angle in Eq.(\ref{1+pure}) remains unchanged $\varphi_{nP}=\theta_{nP}$ (n=1,2,3).

\subsubsection{Partial waves and the ${}^1D_2-{}^3D_2$ mixing angle of the $2^-$ state}

\begin{figure}
\centering
\includegraphics[width=0.43\textwidth]{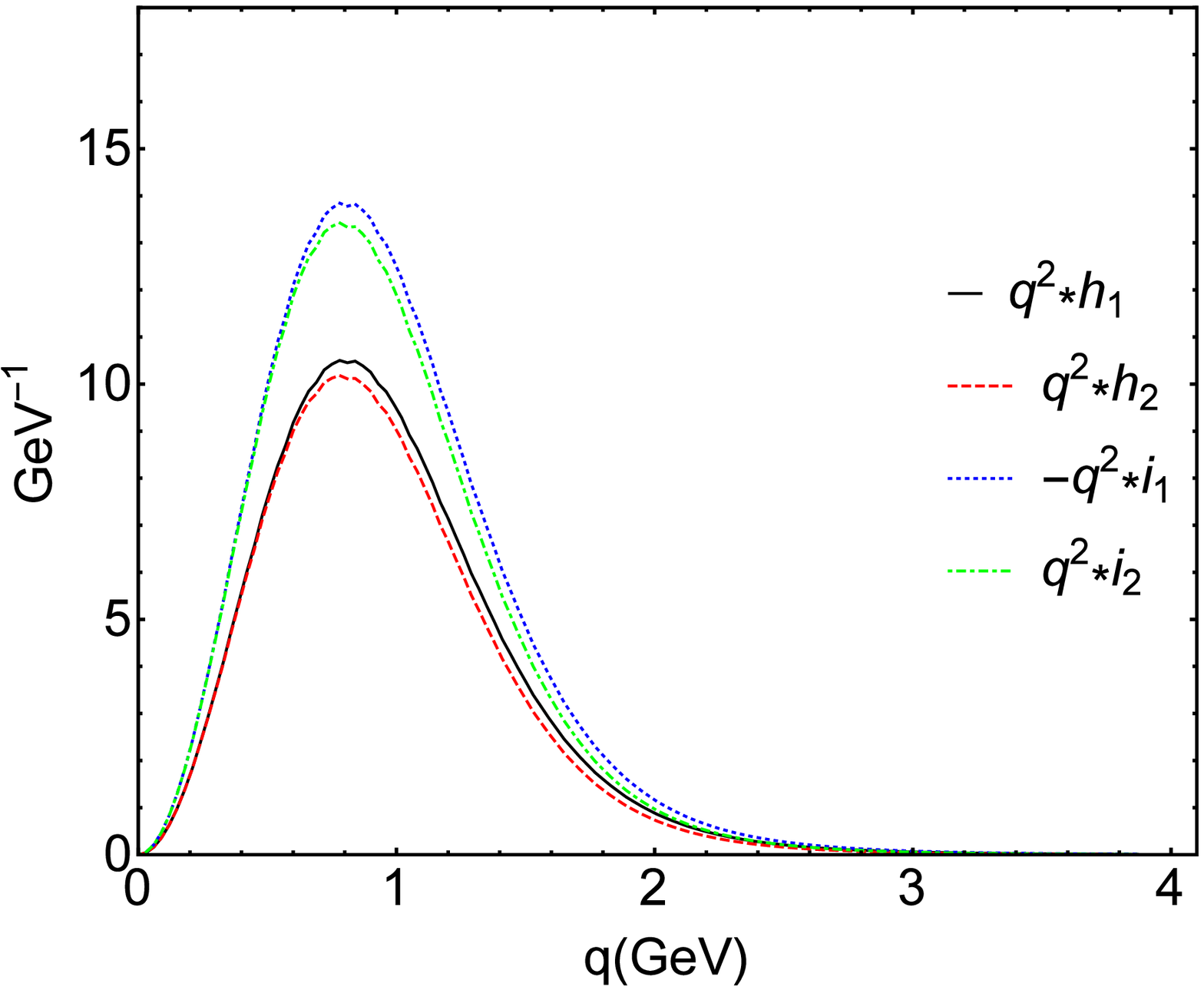}
\includegraphics[width=0.43\textwidth]{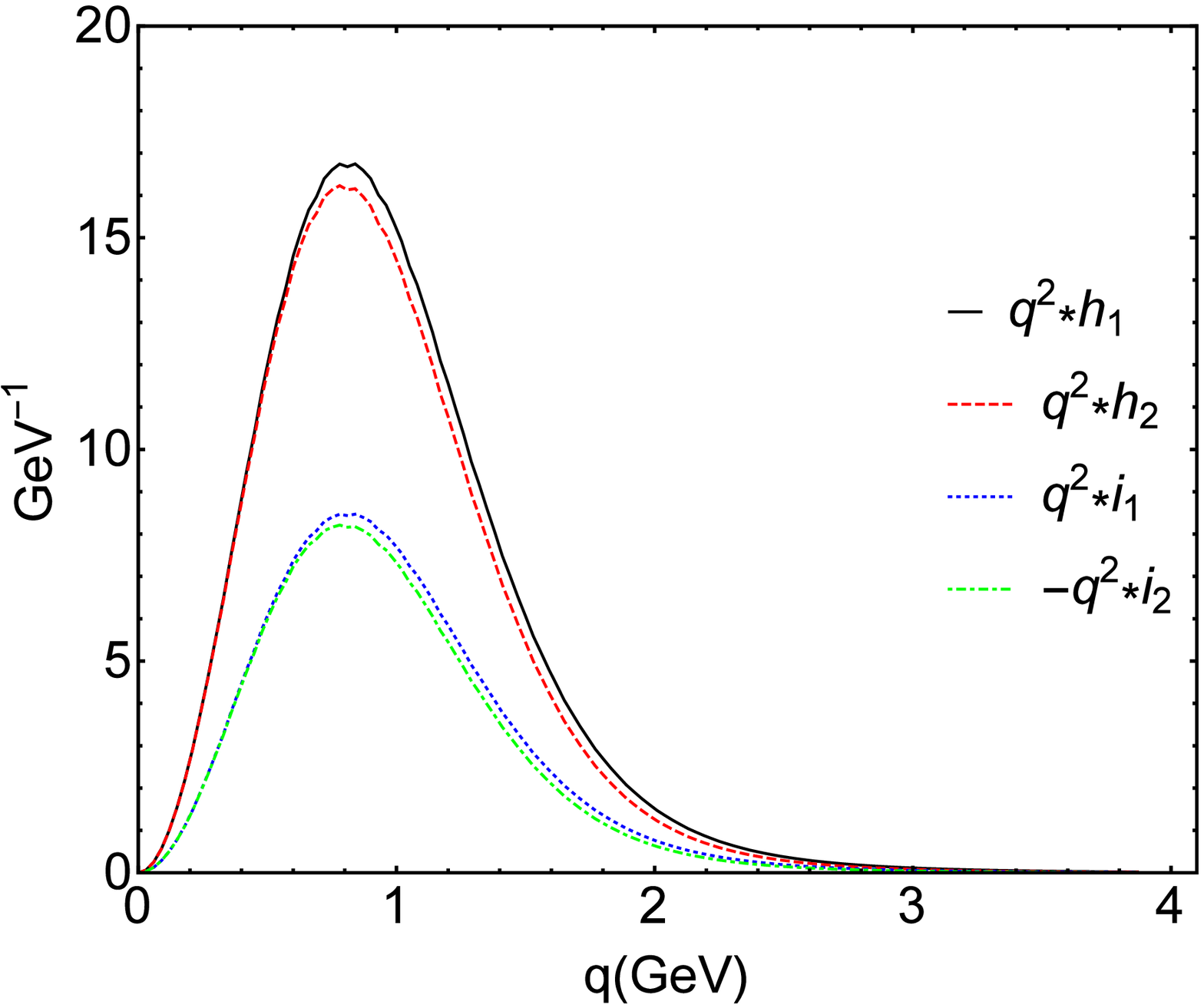}
\caption{The $2^-$ wave functions of the $1^1D_2-1^3D_2$ mixing states $B_{c2}(1D)$ (left) and $B'_{c2}(1D)$ (right). $h_1$ and $h_2$ terms are $^1D_2$ waves; $i_1$ and $i_2$ terms are $^3D_2$ waves.} \label{f2-}
\end{figure}

The solutions of the $2^-$ state also appear in pairs, which is similar to that of the $1^+$ state. Both states of the first pair are $1D$ waves. According to the normalization formula Eq.(\ref{2-nor}), we obtain the following ratios,
$1{}^1D_2 : 1{}^3D_2=0.277:0.723$ and $0.723:0.277$. And the corresponding mixing angle between $1{}^1D_2$ and $1{}^3D_2$ is $\theta_{1D}=-58.2^\circ$ or $31.8^\circ$. The third and forth solutions form a pair, which are both $2D$ states. The following ratios ${2}^1D_2 : {2}^3D_2=0.291:0.709$ and $0.709:0.291$, and the mixing angle $\theta_{2D}=-57.4^\circ$ or $32.6^\circ$ are obtained. For the fifth and sixth solutions which are all $3D$ states, we get ${3}^1D_2 : {3}^3D_2=0.289:0.711$ and $0.711:0.289$, $\theta_{3D}=-57.5^\circ$ or $32.5^\circ$. {As an example, we show the wave functions of the $B_{c2}(1D)$ and $B'_{c2}(1D)$ states in Figure \ref{f2-}.}

We also point out that, the $2^-$ state is not a pure $D$ wave, but includes a small amount of $F$ wave. The $h_1$, $h_2$, $i_1$ and $i_2$ terms in Eq.(\ref{2-wave}) are $D$ waves, but $h_3$, $h_4$, $i_3$ and $i_4$ terms are $F$ waves. The ratios we get are as follows: $D:F=1:0.106$ for the first and second solutions, which means they are two $1D$ dominant states with small amount of $F$ wave; $D:F=1:0.108$ for two $2D$ states and $D:F=1:0.114$ for two $3D$ states. If we ignore the small $F$ wave contribution, similar to the $1^+$ case, the mixing angle in Eq.(\ref{2-pure}) remains unchanged $\varphi_{nD}=\theta_{nD}$ (n=1,2,3).

The mixings in the $1^+$ and $2^-$ states are different from those in the $1^-$, $2^+$ and $3^-$ states. For example, our results show that the partial waves $(n+1)^3S_1$ and $n^3D_1$ in the $1^-$ states are irrelevant, they do not appear in pairs, and do not share the same mixing angle. On the contrary, each pair of the $nP$ $1^+$  states or $nD$ $2^-$ states are related, and there are mixings between $n^1P_1-n^3P_1$ or $n{}^1D_2-n{}^3D_2$. They share the same mixing angle. So the mixings in the $1^+$ and $2^-$ states are consistent with those which are generally considered in the literature.

\subsection{conclusion}

In this article, by solving the instantaneous Bethe-Salpeter equation,  we studied the mass spectrum and wave functions of the $B_c$ system. We also calculated the partial waves in each $J^P$ state, and found that they were not pure $S$, $P$ or $D$ waves, but all contained other components. The method we used is to solve the relativistic Salpeter equation to get the partial waves of a state. This is different from other methods which use the interaction potential, or fit the experimental data of the decay or production processes.
For the $1^+$ and $2^-$ states, we get the ${}^1P_1-{}^3P_1$ and ${}^1D_2-{}^3D_2$ mixings, respectively, which are the same as those in the literature.

With regard to the well-known $S-D$ mixing in the $1^-$ state, we did not get the $(n+1)^3S_1 - n^3D_1$ type mixing provided by the non-relativistic methods. Our results show that in the $S$ wave dominant state, the proportions of $P$ and $D$ partial waves are small and can be ignored, while in the $D$ wave dominant state, the $S$ and $P$ wave components are all very large. We also get similar conclusions in the $2^+$ and $3^-$ states.

%%%%%%%%%%%%%%%%%%%%%%%%%%%%%%%%%%%%%%%%%%%%%%%%%%%%%%%%%%%%%%%%%%%%%
\vspace{0.7cm} {\bf Acknowledgments}

This work was supported in part by the National Natural Science Foundation of China (NSFC) under the Grants Nos. 12075073, 12005169, 12075301, 11821505 and 12047503, the Natural Science Foundation of Hebei province under the Grant No. A2021201009, and Natural Science Basic Research Program of Shaanxi under the Grant No. 2021JQ-074.
%%%%%%%%%%%%%%%%%%%%%%%%%%%%%%%%%%%%%%%%%%%%%%%%%%%%%%%%%%%%%%%%%%%%

\begin{appendix}
\section{Introduction of the Bethe-Salpeter equation and the Salpeter equation}\label{appendix}

The Bethe-Salpeter equation (BSE) \cite{BS} is a relativistic dynamic equation which describes the two-body bound state. For a meson with quark 1 and antiquark 2, the BSE is written as
\begin{equation}
\label{BS}
\chi_{_P}(q)
=iS(p_1)\int{\frac{d^4k}{(2\pi)^4}V(P,k,q)
\chi_{_P}(k)}S(-p_2),
\end{equation}
where $\chi_{_P}(q)$ is the relativistic wave function for the meson with total momentum $P$ and internal relative momentum $q$;
$S(p_1)$ and $S(-p_2)$ are the propagators of quark and antiquark, respectively; $V(P,k,q)$ is the interaction kernel.
The momenta of quark and antiquark are expressed by
$p_1=\frac{m_1}{m_1+m_2}P+q$ and $
p_2=\frac{m_2}{m_1+m_2}P-q$, respectively.

Since the BSE is hard to solve, we choose the instantaneous approximation, which is suitable for a double heavy meson.
With such approximation, the kernel $V(P,q,k)$ can be written as $V(q_{_{\bot}},~k_{_{\bot}})$, where $$q_{_{\perp}}= q-q_{_P}\frac{P}{M},~~~~
q_{_P}=\frac{P\cdot q}{M}.$$ In the rest frame of the meson, we have {$P=(M,0)$}, $q_{_{\perp}}=(0,\vec q)$, and $q_{_P}=q_{_0}$.

For simplicity, we define the three dimensional relativistic wave function $\varphi(q_{_{\perp}})$ and the shorthand symbol $\eta_{_P}(q_{_{\perp}})$ as
$$\varphi(q_{_{\perp}}) \equiv i \int{\frac{dq_{_P}}{2\pi} \chi_{_P}(q)},
\qquad
\eta_{_P}(q_{_{\perp}}) \equiv \int{\frac{dk^3_{_{\perp}}}{(2\pi)^3} V(k_{_{\perp}},q_{_{\perp}})\varphi(k_{_{\perp}}) }.$$
Then the BSE is changed to
\begin{equation}
\label{BS1}
\chi_{_P}(q)=S(p_1)\eta_{_P}(q_{_{\perp}})S(-p_2),
\end{equation}
where in the $B_c$ system, the leading order propagators $S(p_1)={i}/{({\not\!p}_1-m_1)}$ and $S(-p_2)={i}/{(-{\not\!p}_2-m_2)}$ are good choices. They are written as
$$
iS(p_{1}) = \frac{\Lambda^+(p_{1_{\perp}})}{p_{1_P}-\omega_{1}+i\epsilon} + \frac{\Lambda^-(p_{1_{\perp}})}{p_{1_P}+\omega_{1}-i\epsilon},$$$$
-iS(-p_{2}) = \frac{\Lambda^+(-p_{2_{\perp}})}{-p_{2_P}-\omega_{2}+i\epsilon} + \frac{\Lambda^-(-p_{2_{\perp}})}{-p_{2_P}+\omega_{2}-i\epsilon},
$$
where we have defined $\omega_{1}=\sqrt{m_{1}^2-p^2_{1_{\perp}}}$ and $\omega_{2}=\sqrt{m_{2}^2-p_{2_{\perp}}^2}$. The expressions of the projection operators are
$$\Lambda^\pm(p_{1_{\perp}})=\frac{1}{2\omega_{1}}\Big[\frac{\slashed P}{M}\omega_{1} \pm(m_{1}+\slashed p_{1_{\perp}})\Big],
$$$$\Lambda^\pm(-p_{2_{\perp}})=\frac{1}{2\omega_{2}}\Big[\frac{\slashed P}{M}\omega_{2} \pm(-m_{2}+\slashed p_{2_{\perp}})\Big],
$$
{which satisfy the} relations:
$\Lambda^+(p_{1_{\perp}})+\Lambda^-(p_{1_{\perp}})=\frac{\slashed P}{M}$, $\Lambda^\pm(p_{1_{\perp}})\frac{\slashed P}{M}\Lambda^\pm(p_{1_{\perp}})=\Lambda^\pm(p_{1_{\perp}})$, $\Lambda^\pm(p_{1_{\perp}})\frac{\slashed P}{M}\Lambda^\mp(p_{1_{\perp}})=0$, and the similar expressions for the case of antiquark.

Using the contour integral method, we can integrate out $q_{_P}$ on both sides of Eq.(\ref{BS1}), and obtain the Salpeter equation \cite{Sal}
\begin{equation}
\varphi(q_{_{\perp}})=\frac{
\Lambda^{+}(p_{1_{\perp}})\eta_{_P}(q_{_{\perp}})\Lambda^{+}(-p_{2_{\perp}})}
{(M-\omega_{1}-\omega_{2})}- \frac{
\Lambda^{-}(p_{1_{\perp}})\eta_{_P}(q_{_{\perp}})\Lambda^{-}(-p_{2_{\perp}})}
{(M+\omega_{1}+\omega_{2})}\;.
\end{equation}

With the definitions
\begin{equation}\label{project}
\varphi^{\pm\pm}=
\Lambda^{\pm}(p_{1_{\perp}})
\frac{\not\!{P}}{M}\varphi \frac{\not\!{P}}{M}
\Lambda^{{\pm}}(-p_{2_{\perp}})\;,
\end{equation}
the wave function can be divided into four parts
\begin{equation}
\varphi(q_{_{\perp}})=\varphi^{++}(q_{_{\perp}})+
\varphi^{+-}(q_{_{\perp}})+\varphi^{-+}(q_{_{\perp}})
+\varphi^{--}(q_{_{\perp}}),
\end{equation}
where $\varphi^{++}(q_{_{\perp}})$ is called positive wave function, and $\varphi^{--}(q_{_{\perp}})$ negative wave function.
Using Eq.(\ref{project}) and the relations of projection operators, we can rewritten the Salpeter equation as four independent equations,
\begin{equation}\label{posi}
\varphi^{++}(q_{_{\perp}})=\frac{
\Lambda^{+}(p_{1_{\perp}})\eta(q_{_{\perp}})\Lambda^{+}(-p_{2_{\perp}})}{(M-\omega_{1}-\omega_{2})}\;,
\end{equation}
\begin{equation}\label{nega}\varphi^{--}(q_{_{\perp}})=-\frac{
\Lambda^{-}(p_{1_{\perp}})\eta(q_{_{\perp}})\Lambda^{-}(-p_{2_{\perp}})}{(M+\omega_{1}+\omega_{2})}\;,
\end{equation}
\begin{equation}
\varphi^{+-}(q_{_{\perp}})=\varphi^{-+}(q_{_{\perp}})=0\;.
\label{const}
\end{equation}

In the main range of $q_{_{\perp}}$, $M+\omega_{1}+\omega_{2}$ is much larger than $M-\omega_{1}-\omega_{2}$, which results in a large positive wave function $\varphi^{++}(q_{_{\perp}})$ and a very small negative wave function $\varphi^{--}(q_{_{\perp}})$. So {one may think Eq.~(A7) and (A8) can be neglected, and it is enough to solve} Eq. (\ref{posi}) only. However, we point out that this will lost the benefit of relativistic Salpeter equation, because {solving} Eq. (\ref{posi}) only,  one just gets the wave function with one parameter, and the relativistic wave function will not be obtained. If one wants to get a relativistic wave function, all four equations should be considered.
\end{appendix}

\end{document}